\documentclass[a4paper,usenatbib]{mnras}
\usepackage{graphicx}
\usepackage{epstopdf}
\usepackage{txfonts}
\usepackage{natbib}
\usepackage{url}
\usepackage{multirow,bigdelim,multicol}
\usepackage{placeins}
\usepackage{longtable}
\usepackage{threeparttable}
\usepackage{ulem}
\usepackage{color}
\usepackage{bm}
\usepackage{array}

\DeclareRobustCommand{\ion}[2]{\textup{#1\,\textsc{\lowercase{#2}}}}
\newcommand{\HI}{\ion{H}{i}}
\newcommand{\CI}{\ion{C}{i}}
\newcommand{\ClI}{\ion{Cl}{i}}

\def\h2{$\rm H_2$}

\newcommand{\kms}{\ensuremath{{\rm km\,s^{-1}}}}
\newcommand{\qso}{J$\,$0843$+$0221}
\definecolor{orange}{rgb}{1.0,0.5,0.0}

\definecolor{violet}{rgb}{0.53,0.08,1.0}

\title[Very large H$_2$ content in the DLA at z=2.7865]{CO-dark molecular gas at high redshift: very large H$_2$ content and high pressure in a low metallicity damped Lyman-alpha system
}

\author[S. A. Balashev et al.]{S.A.~Balashev\,$^{1}$\thanks{E-mail: {s.balashev@gmail.com}},
    P.~Noterdaeme\,$^{2}$,
    H.~Rahmani\,$^{3,4}$,
	V.V.~Klimenko\,$^{1}$,
	C.~Ledoux\,$^{5}$,
	\and
	P.~Petitjean\,$^{2}$,
	R.~Srianand\,$^{6}$,
	A.V.~Ivanchik\,$^{1}$,
	D.A.~Varshalovich\,$^{1}$
	\\
	$^{1}$Ioffe Institute of RAS, {Polytekhnicheskaya 26}, 194021 Saint-Petersburg, Russia\\
	$^{2}$Institut d'Astrophysique de Paris, CNRS-UPMC, UMR7095, 98bis bd Arago, 75014 Paris, France \\
	$^{3}$Aix Marseille Universit\'e, CNRS, LAM (Laboratoire d'Astrophysique de Marseille) UMR 7326, 13388, Marseille, France \\
	$^{4}$School of Astronomy, Institute for Research in Fundamental Sciences (IPM), P.O. Box 19395-5531, Tehran, Iran \\
	$^{5}$European Southern Observatory, Alonso de C\'ordova 3107, Vitacura, Casilla 19001, Santiago 19, Chile \\
	$^{6}$Inter-University Centre for Astronomy and Astrophysics, Post Bag 4, Ganeshkhind, 411\,007 Pune, India \\
}

\date{Accepted 2017 May 26; in original form 2017 February 17}

\pubyear{2017}

\begin{document}
\label{firstpage}
\pagerange{\pageref{firstpage}--\pageref{lastpage}}
\maketitle
	
\begin{abstract}
  We present a detailed analysis of an H$_2$-rich, extremely strong intervening damped \mbox{Ly-$\alpha$} absorption system (DLA) at $z_{\rm abs}=2.786$ towards the quasar \qso, observed with the Ultraviolet and Visual Echelle Spectrograph on the Very Large Telescope. The total column density of molecular (resp. atomic) hydrogen is $\log N$(H$_2$)=$21.21\pm0.02$  (resp. $\log N$(H\,{\sc i})=$21.82\pm0.11$), making it to be the first case in quasar absorption line studies with H$_2$ column density as high as what is seen in $^{13}$CO-selected clouds in the Milky-Way. 
  
  We find that this system has one of the lowest metallicity detected among H$_2$-bearing DLAs, with $\rm [Zn/H]=-1.52^{+0.08}_{-0.10}$. This can be the reason for the marked differences compared to systems with similar H$_2$ column densities in the local Universe: $(i)$ the kinetic temperature, $T\sim$120~K, derived from the $J=0,1$ H$_2$ rotational levels is at least twice higher than expected; $(ii)$ there is little dust extinction with $A_{\rm V} < 0.1$; $(iii)$ no CO molecules are detected, putting a constraint on the $X_{\rm CO}$ factor $X_{\rm CO}> 2\times 10^{23} $ cm$^{-2}$/(km/s\,K), in the very low metallicity gas. Low CO and high H$_2$ contents indicate that this system represents "CO-dark/faint" gas.
  
  We investigate the physical conditions in the H$_2$-bearing gas using the fine-structure levels of \ion{C}{I}, \ion{C}{ii}, \ion{Si}{ii} and the rotational levels of HD and H$_2$. We find the number density to be about $n \sim 260-380$\,cm$^{-3}$, implying a high thermal pressure of $3-5 \times 10^4$\,cm$^{-3}$\,K. We further identify a trend of increasing pressure with increasing total hydrogen column density. This independently supports the suggestion that extremely strong DLAs (with $\log$\,N(H) $\sim 22$) probe high-z galaxies at low impact parameters.

\end{abstract}


\section{Introduction}
\label{introduction}
\noindent

The processes of star formation, galaxy evolution and AGN activities are directly connected with the evolution of different phases of the neutral interstellar medium (ISM). Recent kpc-scale observations of nearby normal galaxies (non-starbursts) have shown that the surface star formation rate ($\Sigma_{\rm SFR}$) in their inner parts correlates approximately linearly with the surface density of molecular hydrogen ($\Sigma_{{\rm H}_2}$), but not with that of atomic gas \citep[see][]{Bigiel2008, Leroy2013}.  In other words, the star formation in { inner parts of} nearby galaxies is closely associated with the molecular phase of the ISM. While such a tight correlation was found for $\Sigma_{\rm SFR} \gtrsim 10^{-3}\, \rm M_{\odot}\,yr^{-1}\,kpc^{-2}$ or $\rm \Sigma_{H_2} > 1\, M_{\odot}\, pc^{-2}$ \citep{Schruba2011}, sensitive measurements in environments with lower star formation rate (SFR) activity, such as dwarf irregular galaxies or the outer parts of the normal nearby galaxies, show that SFR still correlates with \HI\ \citep{Bigiel2010a, Bigiel2010b, Roychowdhury2014}. This could be the consequence of a H$_2$--\HI\ correlation at low gas surface density \citep{Schruba2011}. 

The correlation between star formation and H$_2$ gas is also indirectly supported by the physical conditions
of the latter, which shows that H$_2$ is usually associated with the cold and dense phase of the neutral ISM
that is believed to provide the raw material for star formation. In the cold ISM, it is expected that most of hydrogen will eventually be in molecular form, since the latter is energetically favoured. However, this does not tell whether molecular hydrogen is an essential ingredient for star formation or only a tracer of the cold and dense ISM \citep{Krumholz2011, Krumholz2012, Glover2012a, Michaowski2015, Glover2016}. The basis for
this question is that molecular hydrogen is not major coolant of the cold ISM even at low metallicities. However, the cooling of the gas through atomic emission lines depends on the metal abundance, to which the formation of H$_2$ is also linked, via the correlation between gas-phase and dust-phase metal abundances. In addition, H$_2$ is vital for other complex molecules to form, which can cool dense regions of the ISM and eventually initiate the collapse of the cloud, leading to the formation of stars. 

It is not yet established whether the relation between star formation and H$_2$ surface/column density is valid in high-redshift galaxies, for which the metallicities are expected to be much lower on average \citep[e.g.][]{Rafelski2012}. For example, recent works suggest that at low metallicities star formation can occur before the \HI\ converts into H$_2$ \citep{Krumholz2011, Hu2016}. In addition, $\Sigma_{\rm SFR}$ seems to remain well correlated with \HI\ in dwarf irregular galaxies with metallicities $\sim 0.1 \times Z_{\odot}$ \citep{Roychowdhury2014}. The analysis of damped Ly-$\alpha$ systems (DLAs) detected in the spectra of quasar and gamma-ray bursts (GRB) afterglows provides a unique method to probe the atomic and molecular hydrogen at high redshifts \citep[e.g.][]{Noterdaeme2015}. However, the relation with star-formation remains difficult to establish due to the paucity of direct detections of the associated galaxies, although global scaling relations (between e.g. SFR, metallicity and impact parameter) start to be revealed (Krogager et al. 2017). The measured SFR in the overall population of intervening DLAs appears to be very low \citep{Fumagalli2015, Rafelski2016}, while DLAs associated with GRB afterglows span a range of SFRs \citep{Toy2016}. 

Finally, there is no unambiguous way to convert measurement of column densities towards one particular line of sight to the overall properties of the galaxy. In spite of that, it is known that H$_2$ is less abundant in DLAs at high redshifts (such DLAs typically show lower metallicity), i.e. in the diffuse ISM of high-redshift galaxies: only a small fraction of high-$z$ DLAs ($<10$ per cent; \citealt{Noterdaeme2008, Balashev2014, Jorgenson2014,  Noterdaeme2015b}) show associated H$_2$. This contrasts with observations in the local Universe, where the fraction of \HI\ absorbers with associated H$_2$ is $\sim90\%$ for our Galaxy \citep{Wakker2006} and $\sim70\%$ for Magellanic Clouds \citep{Tumlinson2002} with molecular fraction $\rm f > 10^{-6}$. 

It is hard to attribute the difference in the H$_2$ detection rate between high redshifts and the local Universe to a pure observational effect (obscuration of the background quasar by dust associated with H$_2$-bearing clouds), since it was found that in the  majority of DLAs at high redshifts (z~$>1.8$), the dust extinction is negligible \citep[e.g][]{Murphy2016}, although some highly reddened
cases do exist \citep[e.g.][]{Srianand2008b, Noterdaeme2009a, Ma2015, Krogager2016, Noterdaeme2017}, in which cold gas is found. Therefore the low incidence rate of H$_2$ is most probably connected with the small covering factor of H$_2$-bearing cold ISM at high redshifts.

The small incidence rate of H$_2$ absorption systems in high redshift DLAs also implies that only about 30 H$_2$ absorption systems at high redshifts have been identified and therefore the cold ISM remains least studied. Blind searches for H$_2$ absorption systems are quite ineffective because the detection of molecular absorption usually requires high-resolution (or intermediate in the case of very saturated H$_2$ absorption system) and high-S/N spectrum to overcome the effect of contamination by the Ly$\alpha$ forest. In addition, among the known high-$z$ H$_2$-bearing DLAs, only few have high H$_2$ column density, $\log N \gtrsim 19$\footnote{Here and after, the column densities are quoted in cm$^{-2}$}, and high molecular fraction, $f\gtrsim0.1$.

Studies of low-metallicity H$_2$-bearing systems can be used to estimate the fraction of CO-dark molecular gas, which is not seen in CO emission \citep{Wolfire2010}. A recent work indicates that such gas is important for star formation, since it is more susceptible to turbulent compression and gravitational collapse than pure atomic gas \citep{Glover2016}. It was measured that a large fraction of the molecular gas can be in CO-dark phase: observations of \ion{C}{II} in the Milky Way (MW) show that, on an average, the CO-dark gas can contain as much as $\sim$30\% of the total H$_2$ content \citep{Pineda2013}. This fraction is expected to increase with decreasing metallicity \citep{Glover2012b}. Indeed, at low metallicity the relative abundance of CO to H$_2$ is reduced by the reduced abundances of C and O together with a weaker dust extinction.  Observationally, neither the fraction of CO-dark gas nor $X_{\rm CO}$, the conversion factor between CO emission and H$_2$ mass \citep[see the recent review by][]{Bolatto2013} is well constrained in galaxies both at high redshifts and with low metallicity \citep[][and references therein]{Shi2016}. Any measurements of these parameters are very valuable, since with the advent of state-of-the-art sub-mm facilities like ALMA, CO can now be detected in normal galaxies at high redshifts.

Several efficient techniques to search for molecular-rich absorption systems at high redshift have recently been proposed. These techniques use different target pre-selections in the Sloan Digital Sky Survey \citep[SDSS;][]{York2000} spectroscopic data base. One technique based on the presence of strong neutral carbon lines in SDSS quasar spectra \citep[see][]{Ledoux2015} was proposed to select
high-metallicity CO/H$_2$-bearing systems. This led to the first studies of the translucent ISM at high redshifts \citep[e.g.][]{Srianand2008, Noterdaeme2009a, Noterdaeme2010}. It was also shown that the 2175\AA\ bump can be detected in SDSS spectra \citep[e.g.][]{Wang2004, Srianand2008b, Jiang2011} and could potentially be used to preselect molecular-rich absorbers. Indeed the follow-up studies \citep[e.g.][]{Ma2015, Noterdaeme2017} showed that both \ion{C}{i} and 2175\AA\ bump selection techniques probe molecular bearing
systems with high metallicities, up to super-solar. These are at the high end of the metallicity distribution of high-redshift DLAs and could therefore probe gas with chemical enrichment closer to what is seen in the local Universe than in typical high-z galaxies.

Another technique was proposed by \citet{Balashev2014} to directly search for H$_2$ absorption systems at $z>2.3$ in SDSS quasar spectra. In short, the technique uses a template matching routine to automatically detect damped Lyman and Werner H$_2$ lines using the SDSS DLA catalogue \citep{Noterdaeme2012}.  A confidence is attributed to each candidate based on a resampling method and using control samples. While only about $\lesssim 3$ per cent of SDSS spectra are suitable to search for H$_2$, and only very high column density H$_2$ absorption systems (with $\log N \gtrsim 19.5$) can be detected, it is worth mentioning that this technique does not depend on metal tracers, and should therefore probe high $N$(H$_2$) over a wide range of metallicities. 

To date, we have detected $\sim100$ reliable H$_2$ candidates using SDSS DR12 and started follow-up observations with the Very Large Telescope (VLT). All of the eight candidates followed-up so far have been confirmed as strong H$_2$ systems.

In this paper, we present the study of the system with strongest H$_2$ absorption in our sample, at $z_{\rm abs}=2.786$ towards the quasar SDSS J\,084312.72$+$022117.28  (hereafter \qso) and postpone the study of the overall sample to a future work. H$_2$ lines towards \qso\ are so saturated that they were initially
identified as multiple DLA systems in the automatic DLA search by \citet{Noterdaeme2012}. We obtained high-resolution spectroscopic observations of the system using the Ultraviolet and Visual Echelle Spectrograph (UVES) on the VLT. Section~\ref{observations} presents these observations and the data reduction. Section~\ref{analysis} presents the analysis of the system, including metal lines, associated molecules and dust extinction. We estimate the physical conditions in the gas in Section~\ref{conditions} and discuss our findings in Section~\ref{discussion}. We conclude in Section~\ref{conclusion}. 

\section{Observations and data reduction}
\label{observations}

The observations of \qso\ (z$_{\rm em}$=2.92) were carried out between 2014 January and March
using UVES \citep{Dekker2000} under programme 092.A.0345 (PI: D.A.~Varshalovich). The journal of the observations is presented in Table~\ref{table_obs}. We used two standard settings: 390$+$564 (with dichroic 1 and cross-dispersers CD\#2 and CD\#3) and 437$+$760 (with dichroic 2 and cross-dispersers CD\#2 and CD\#4) to cover most of the metal and molecular absorption lines associated with the DLA at $z=2.7865$.  A slit width of 1.0 arcsec and CCD readout with 2$\times$2 binning  were used for all the observations, resulting in the spectral resolution $R \approx$~48\,000.

We reduced the data using UVES Common Pipeline Library (CPL) data reduction pipeline release 6.5.1\footnote{http://www.eso.org/sci/facilities/paranal/instruments/uves/doc/} with the optimal extraction method. To avoid extra rebinning of the pixels we used the final un-rebinned extracted spectrum of each order produced by CPL and binned it into a global grid of uniform pixel size of 2.5 km~$\rm s^{-1}$.  We applied
the CPL wavelength solutions to each order and merged the orders by implementing a weighted mean in the overlapping regions. All the exposures were shifted to the heliocentric-vacuum frame correcting for the motion of the observatory along the quasar line of sight obtained at the exposure mid-point and using
the air-to-vacuum relation from \citet{Edlen1996}. Then we combined all exposures using a weighted mean at each pixel. The standard deviation around the continuum level in our final combined spectrum is in agreement, within a few percent, with our error spectrum. The final spectrum covers the observed wavelength range of 328--947 nm with gaps of $\sim$ 7 and 11 nm centred at 564 and 760 nm, respectively.

\begin{table*}
	\begin{center}
     \begin{tabular}{ccccccc}
            \hline
            No. & Date & Starting time (UT) & Setting & Exposure (s) & Airmass & Seeing (arcsec)\\
            \hline
            1 & 07.01.2014$\dagger$ & 07:19:05 & 437+760 & 4800 & 1.162$\to$1.377 & 1.00$\to$0.78 \\
            2 & 28.01.2014 & 03:06:44 & 437+760 & 4800  & 1.265$\to$1.131 & 0.83$\to$1.10 \\
            3 & 24.02.2014 & 01:07:58 & 390+564 & 4800 & 1.306$\to$1.141  & 0.97$\to$0.92 \\
            4 & 26.02.2014 & 01:34:21 & 437+760 & 4800 & 1.211$\to$1.122  & 0.81$\to$1.00 \\
            5 & 26.02.2014 & 02:58:31 & 437+760 & 4800 & 1.121$\to$1.188  & 0.98$\to$1.24 \\
            6 & 27.02.2014 & 01:32:14 & 390+564 & 4800 & 1.207$\to$1.122  & 0.99$\to$0.96 \\
            7 & 27.02.2014 & 02:55:32 & 390+564 & 4800 & 1.121$\to$1.190  & 0.94$\to$0.74 \\
            8 & 27.02.2014 & 04:16:46 & 390+564 & 4800 & 1.192$\to$1.466  & 0.69$\to$0.75 \\
            9 & 02.03.2014 & 01:02:50 & 437+760 & 4800 & 1.249$\to$1.127  & 0.69$\to$0.84 \\
            \hline
            Total & \multicolumn{3}{c}{} & 43200 \\
            \hline
     \end{tabular}
     \caption{Journal of the observations
     	$\dagger$ For this exposure calibration files were not attached}
     \label{table_obs}
     \end{center}
\end{table*}

\section{Analysis}
\label{analysis}
We analysed absorption lines from the DLA using a code that simultaneously fits the absorption profiles of selected species using Voigt profile convolved with the instrumental response function. We used $\chi^2$ likelihood function, which assumes the probability distribution function of uncertainties in the spectrum to be Gaussian. The shape of the likelihood function is defined by the Monte Carlo Markov chain approach with implementation of the affine-invariant ensemble sampler \citep{Goodman2010}. Such a technique ensures that we confidently find the global maximum in the many parametric space and provides reliable estimates of statistical errors on the parameters.  Furthermore, the correlation between different parameters can be easily traced by this method. We cross-checked our results for several species with an independent analysis obtained from \textsc{vpfit}\footnote{http://www.ast.cam.ac.uk/$\sim$rfc/vpfit.html} and found a good match between the two approaches. All the errors quoted here are given using 63.8\% quantile interval, which formally corresponds to 1$\sigma$ interval for normal distribution\footnote{In many cases, the derived likelihood function is actually asymmetric, resulting in asymmetric error bars.}.  In each figure presenting a fit to absorption lines (except Figs~\ref{HIprofile} and Fig.~\ref{H2J01}), we grey out pixels of spectrum not used to constrain the fit and show the residuals (data minus model, divided by uncertainty).

\subsection{Neutral hydrogen}
\label{HIcontent}
\begin{figure}
	\begin{center}
		\includegraphics[clip=,width=0.95\hsize]{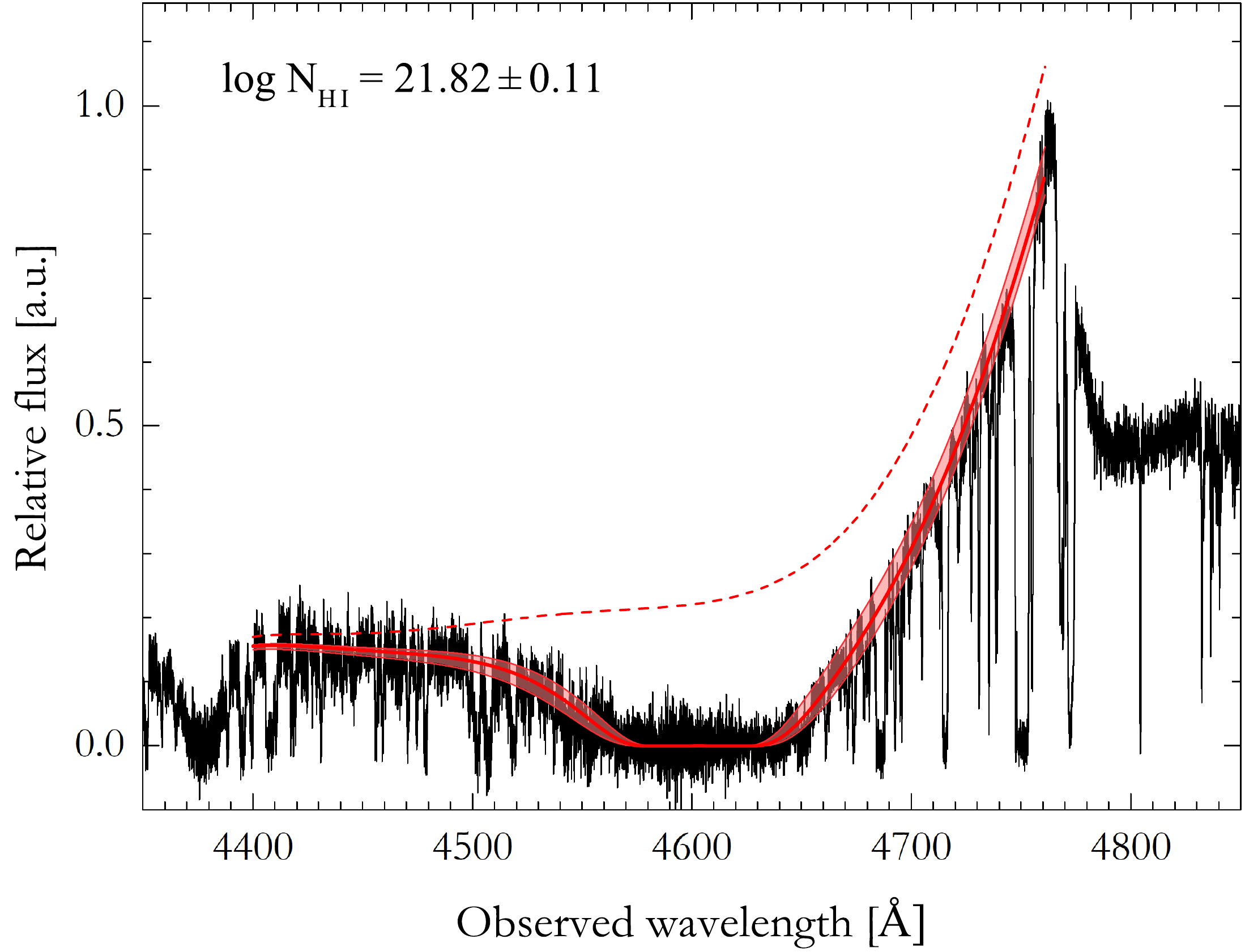}
		\caption{A portion of our UVES spectrum around the DLA at $z=2.786$. The solid red line represents the fitted Ly-$\alpha$ profile { with associated uncertainty as shaded region}.
        The red dashed line shows the reconstructed continuum, which was fitted together with the \HI\ line profile. { The measured \HI\ column density is $\log N$(\HI)=$21.82\pm0.11$, i.e. this DLA corresponds to a sub-class of DLA known as extremely strong DLAs \citep[ESDLAs; ][]{Noterdaeme2014}.}
    	} 
		\label{HIprofile}
	\end{center}
\end{figure}

We measured the \HI\ column density by fitting the damped Ly-$\alpha$ absorption line with a Voigt profile. Higher Lyman series of the \HI\ absorption lines are severely blended with the damped absorptions produced by associated H$_2$. Because the Ly-$\alpha$ line is strongly damped and located near the Ly-$\alpha$ emission line of \qso\ ($\delta v \sim$ 10000 km\,s$^{-1}$), we simultaneously fitted the continuum (using first seven Chebyshev polynomials) and the DLA profile. We obtained $\log N$(\HI)~$\sim 21.82\pm0.11$ at the best-fitted redshift, z$_{\rm DLA}$=$2.78654\pm0.00015$ (see Fig.~\ref{HIprofile}), confirming the value previously derived from SDSS data \citep[$\log N$(\HI)~=~21.8;][]{Noterdaeme2014}.

\subsection{Neutral chlorine as a proxy to the velocity structure of H$_2$}
\label{ClI}
The strong saturation of H$_2$ absorption lines prevents us from deriving the velocity structure
of H$_2$ in its first five rotational levels. Several components can be distinguished in the $J\ge5$ levels but the velocity structure cannot be extracted unambiguously because of the relatively low S/N of the spectrum. Therefore we assume the velocity structure of the H$_2$ to be the same as that of \ClI, which is known to be a good tracer of H$_2$-bearing gas \citep[][]{Jura1974}, both statistically in our Galaxy \citep[][]{Moomey2012} and at high redshifts \citep{Balashev2015}, but also within a given absorbing cloud \citep{Noterdaeme2017}.

\begin{figure}
	\begin{center}
		\includegraphics[clip=,width=0.95\hsize]{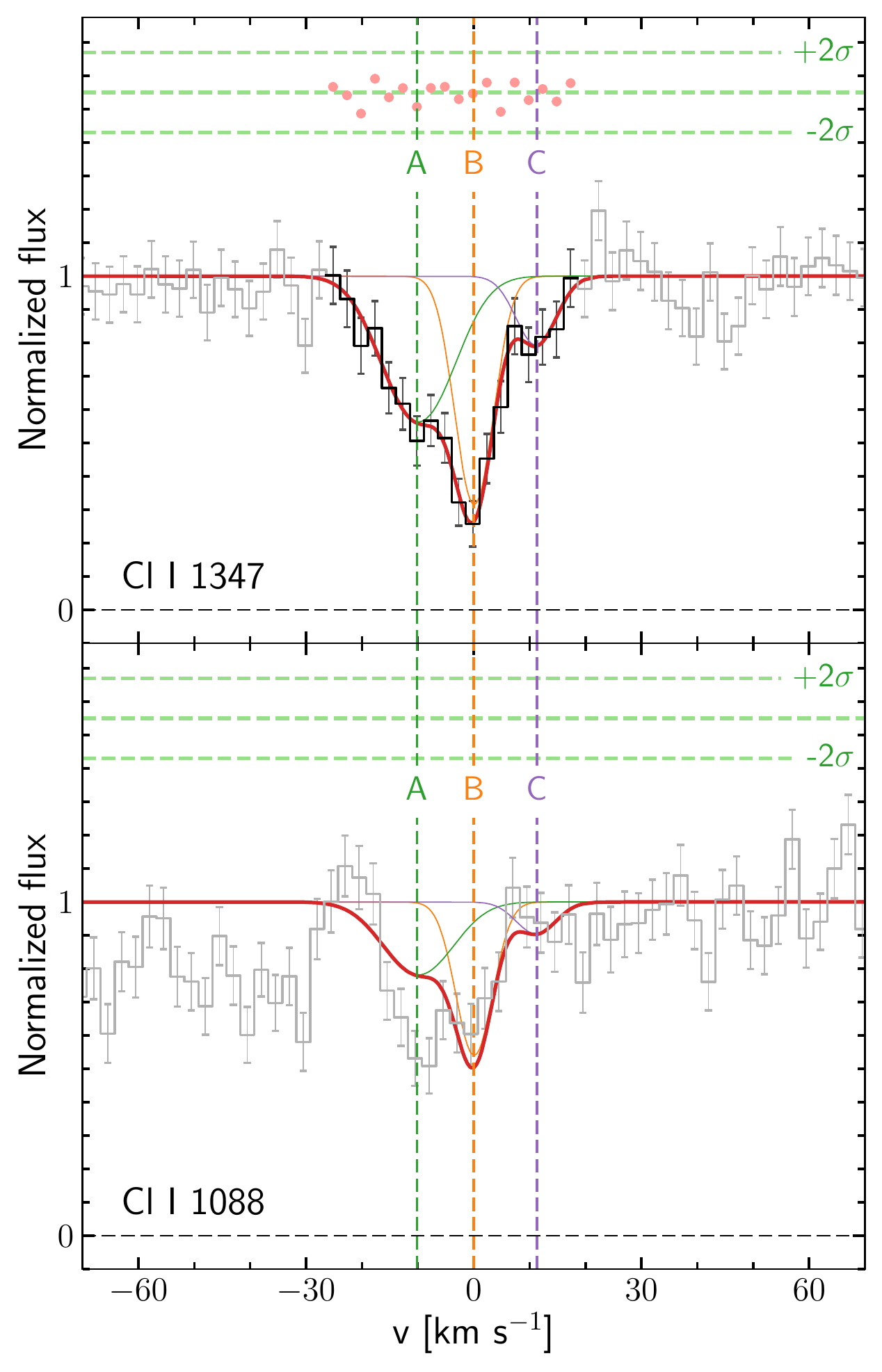}
		\caption{Fit to the \ClI\ absorption lines at z=2.7865. { In this and following figures presenting fits to absorption lines, pixels used (resp. not used) to constrain the fit are represented in black (resp. grey). The synthetic profile is shown in red and the contribution from each component is shown in green, orange and purple for components A, B and C, respectively. The small filled circles at the top of each panel show the residuals and the $\pm 2\,\sigma$ error levels are shown by green lines. We used component structure measured in \ClI\ lines to fit H$_2$ and HD lines.}}
		\label{Chlorine}
	\end{center}
\end{figure}

Absorption lines from two neutral chlorine transitions (\ion{Cl}{i}\,$\lambda$1347 and 1088) are detected in two main components at z=2.786582 and 2.786459 (see Fig.~\ref{Chlorine}). A third, weaker component can be distinguished in the red wing of \ion{Cl}{i}\,$\lambda$1347, which is also seen in the line profiles of H$_2$ with high-$J$ and low-ionization metals. In the following we will denote these components as $A$, $B$ and $C$ in order of increasing redshift. The \ion{Cl}{i}\,$\lambda$1088 line is likely to be blended with unrelated absorption from the Ly-$\alpha$ forest since its apparent blue wing ($v < -5$~\kms) has a higher optical depth than the corresponding region of \ion{Cl}{i}\,$\lambda$1347, despite the oscillator strength of the former being smaller ($f$=0.081) than that of the latter \citep[$f$=0.153; ][]{Schectman1993}. Therefore we excluded \ion{Cl}{i}\,$\lambda$1088 from the fit. The results of the fit are given in Table~\ref{fit_res}. 

\subsection{Molecular hydrogen}
\label{H2}

Molecular hydrogen is detected in the first 10 rotational levels of the ground ($\nu=0$) vibrational state. This is the first detection of such a large number of excited levels from lower vibrational state of H$_2$ in an intervening high-$z$ DLA on QSO line of sight \citep[there are two detections in DLA associated with GRB afterglows;][]{Sheffer2009, Kruhler2013}. 

\begin{figure*}
	\begin{center}
		\includegraphics[clip=,width=0.95\textwidth]{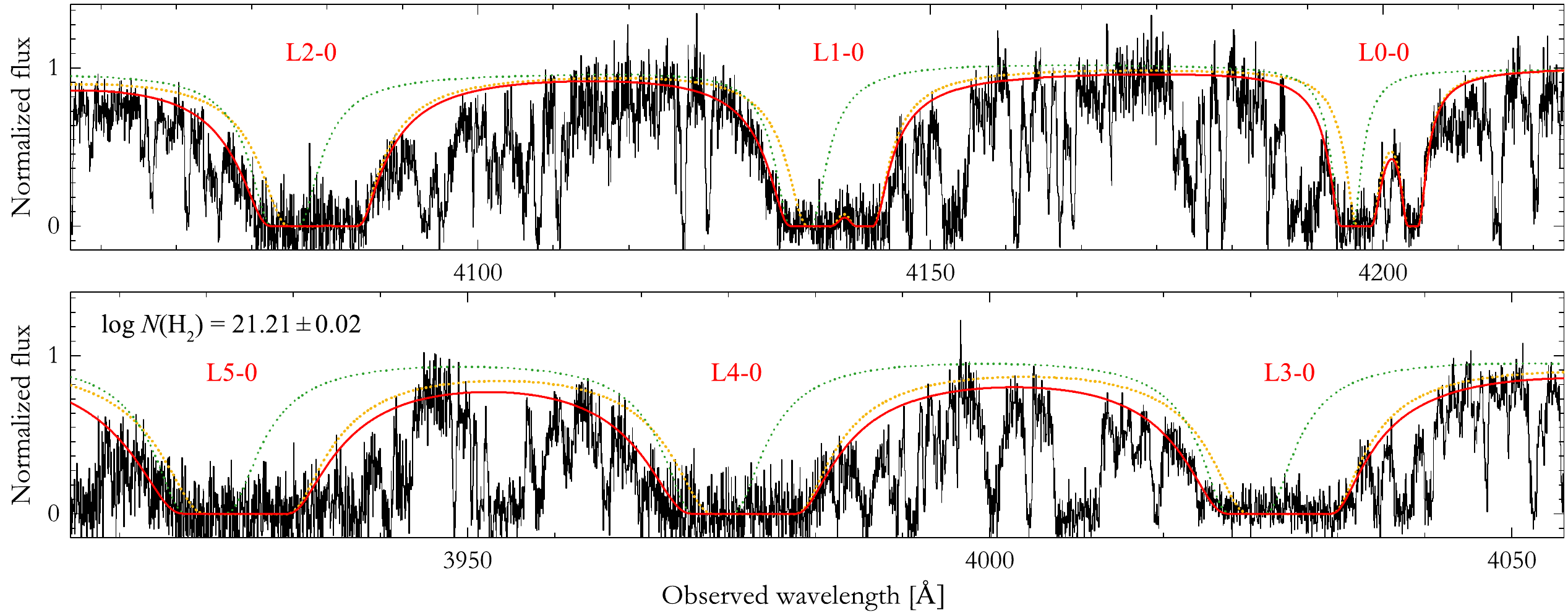}
		\caption{Portion of our UVES spectrum of \qso\ around the region of H$_2$ absorption lines from the DLA at $z=2.786$. The red solid line shows the total fitted profile to the $J=0$ and $J=1$ absorption lines, while green and orange dashed lines represent the contribution of the $J=0$ and $J=1$ levels, respectively. This is the largest H$_2$ column density DLA detected in QSO spectra.
                }
		\label{H2J01}
	\end{center}
\end{figure*}

In all detected Lyman (L) and Werner (W) bands of H$_2$, the $R(0)$, $R(1)$ and $P(1)$ lines are strongly damped and significantly blended with each other (Fig.~\ref{H2J01}). This affects the apparent quasar continuum. We therefore fitted simultaneously the damped absorption lines and the continuum using the first 10
Chebyshev polynomials. This is enough to take into account possible continuum variations due to the blaze
function of our echelle spectrum, which is not flux calibrated. While we have no information on the column
densities in the first two H$_2$ rotational levels for each component individually, the total H$_2$ column density is obtained with high accuracy thanks to the damping wings: we find $\log N$(H$_2$,\,J=0)~=~$20.71\pm0.02$ and $\log N$(H$_2$,\,J=1)~=~$21.06\pm0.02$ at the best-fitted redshift of z=2.786570(14). The best fit to the $J=0$ and $J=1$ H$_2$ absorption lines is shown in Fig.~\ref{H2J01}. We also checked that a fit using three components with velocity structure fixed to that of \ClI\ provides very similar total column densities of H$_2$.

The measured excitation temperature $T_{01}=123^{+9}_{-8}$\,K, is higher than what is usually obtained at high H$_2$ column densities ($\log N$(H$_2)>19$) both in the local Universe and at high redshifts. We will discuss this further in Sect.~\ref{T01}. The fit to $J\ge 2$ rotational levels of H$_2$ is presented in Appendix \ref{H2_appendix} and its results are summarized in Table~\ref{fit_res}.

\subsection{HD molecule}
\label{HD}

\begin{figure}
	\begin{center}
		\includegraphics[clip=,width=0.49\textwidth]{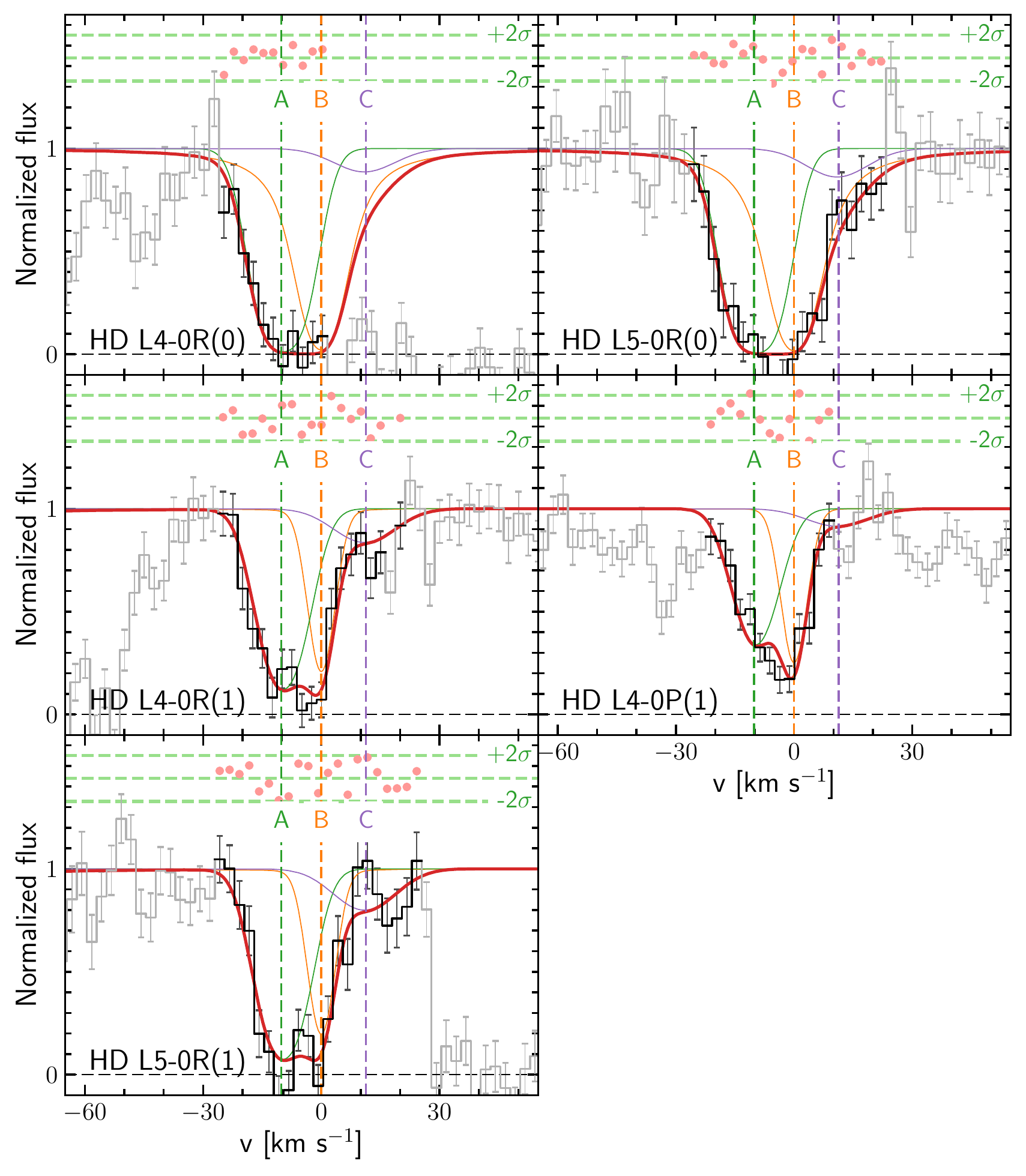}
		\caption{Fit to the HD lines in the $J=0,1$ rotational levels. Lines are as in Fig.~\ref{Chlorine}. { This is the second detection of lines from J=1 rotational levels of HD. Due to saturation of HD lines and low S/N the measured HD column density has large uncertainties.}}
		\label{HDJ01}
	\end{center}
\end{figure}

Absorption lines from deuterated molecular hydrogen (HD) are detected in components $A$ and $B$ in the first two rotational levels. This is only the second detection of HD, J=1 lines at high $z$ \citep[see][for the first detection towards Q$\,$0812+3208]{Balashev2010}. The non-detection of HD in the component $C$ is consistent with low H$_2$ column density in this component. Since HD components are severely blended with each other, we again used fixed redshifts of the components as determined from \ClI. For the typical physical conditions in the cold ISM, the excitation of the first rotational levels of HD from the ground level is expected to be dominated by collisions \citep[see][]{Abgrall1992, Balashev2009}. Therefore we assumed that HD at $J=0$ and $J=1$ levels are co-spatial, and tied together their Doppler parameters. The fit to HD lines is shown in Fig.~\ref{HDJ01} and the parameters are given in Table~\ref{fit_res}. Unfortunately, only two lines from $J=0$ rotational level and three lines from $J=1$ level of HD can be used to constrain the fit. In addition, these lines are in the intermediate regime, leading to large uncertainties on the column densities. We obtain the total column density $\log N$(HD)~=~$17.35^{+0.15}_{-0.34}$.

\subsection{Neutral carbon}
\label{CI}
\begin{figure}
	\begin{center}
		\includegraphics[clip=,width=0.49\textwidth]{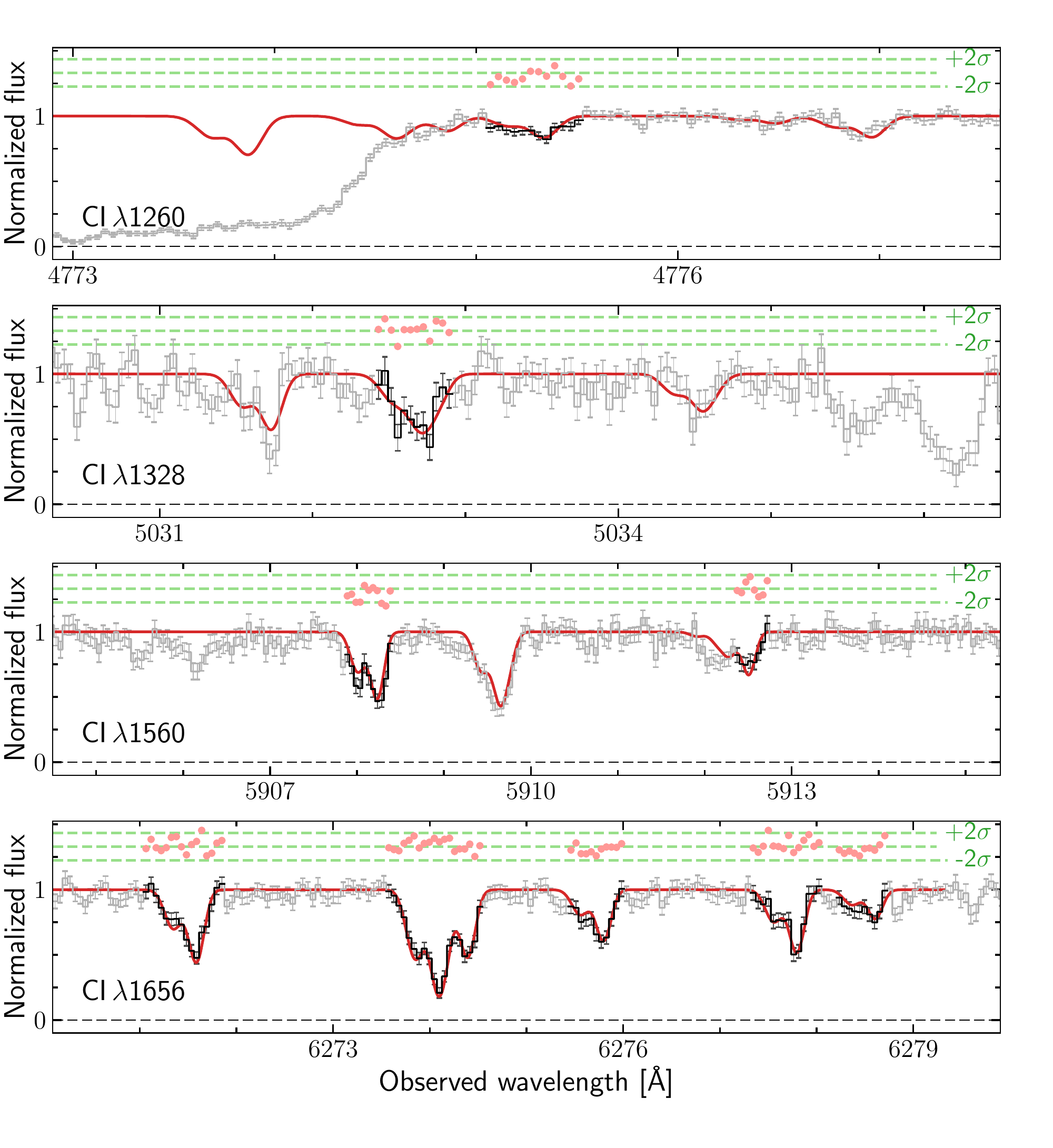}
		\caption{Fit to the \CI\ absorption lines. Lines are as in Fig.~\ref{Chlorine}.}
		\label{CIprofile}
	\end{center}
\end{figure}

We detect \CI\ lines corresponding to transitions from three fine-structure levels of the 2s$^2$2p$^2$ 3P$^0$ \CI\ ground state. As for HD, \CI\ is only detected in components $A$ and $B$. The fit to the \CI\ lines is shown in Fig.~\ref{CIprofile} and the results are given in Table~\ref{fit_res}. The \CI\-to-H$_2$ column density ratio is relatively low, $(6.0\pm0.4)\times10^{-8}$, compared to what has been seen so far in high-$z$ H$_2$-bearing systems \citep[e.g.][]{Srianand2005,Jorgenson2010}. This can indicate an inefficient dust shielding of \CI\ (i.e. higher UV flux and/or low dust content) and/or the low metallicity of the gas (see Sect.~\ref{Metal}) that leads to a low abundance of carbon. 

\subsection{CO molecules}
\label{CO}

Carbon monoxide, which can only survive in the cold H$_2$-bearing gas, has been detected at high redshift in only a few DLA systems. When measurable, these systems show high H$_2$ content \citep{Srianand2008,Noterdaeme2010,Noterdaeme2017}. Here, we do not detect CO absorption lines in spite of the extreme H$_2$ column density. Using several A-X CO bands as well as C-X($0-0$) and E-X($0-0$) CO bands we
derive 3$\sigma$ upper limits on the total CO column density assuming an excitation temperature 10.3$\,$K, which is the cosmic microwave background (CMB) temperature at the absorption redshift $z=2.786$ \citep[see][]{Noterdaeme2011}. We derive $\log N$(CO)~$<$~13.1 and 13.0 for components $A$ and $B$, respectively. 

\subsection{Metal content}
\label{Metal}

Several metal species in different ionization stages (e.g. \ion{O}{i}, \ion{Fe}{ii}, \ion{Si}{iv}) originating from the DLA are detected. While we concentrate here on the low-ionization metals, which provide the necessary information to derive the chemical abundances in the neutral gas, a comparison with highly ionized phase is also provided in Fig.~\ref{ioniz}, for completeness. The main components seen in the profiles of the low-ionization metal lines correspond to the three components identified in \ion{Cl}{i} and H$_2$.

\begin{figure}
	\begin{center}
		\includegraphics[clip=,width=1.0\hsize]{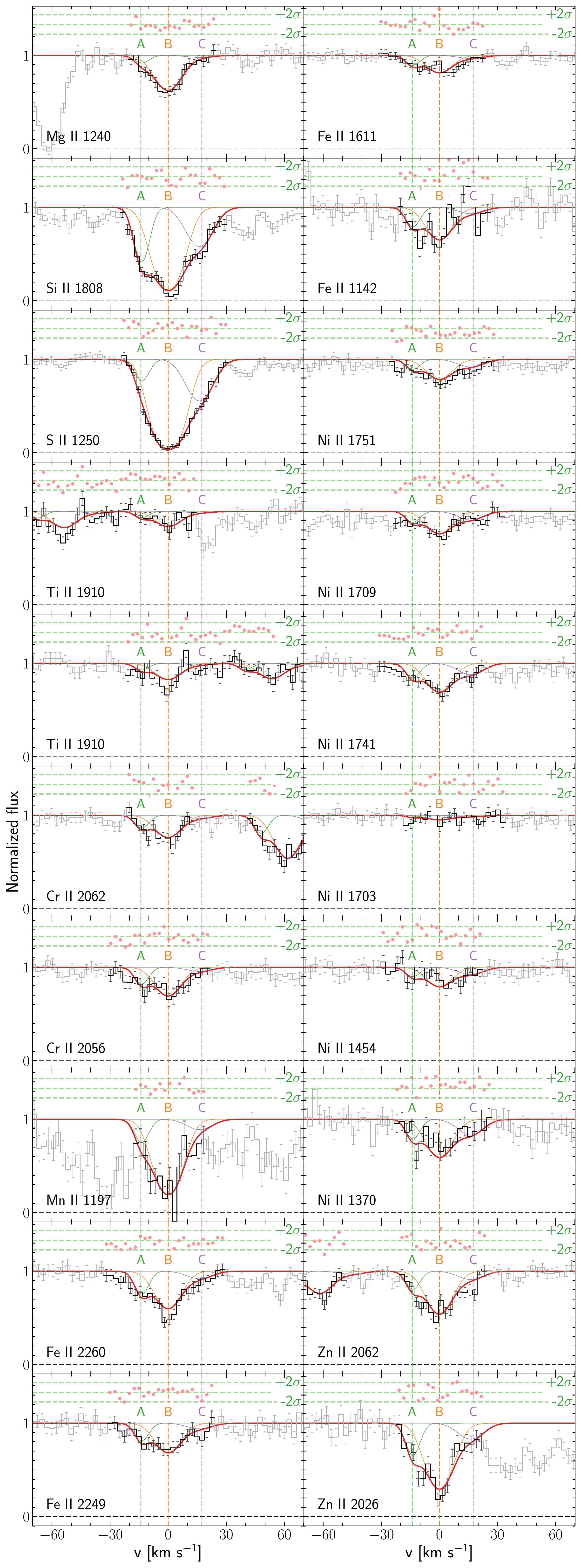}
		\caption{Fit to the low-ionization metal lines. Lines are as in Fig.~\ref{Chlorine}.}
		\label{metals}
	\end{center}
\end{figure}

The redshifts of these components were left free during the fit and found to be in agreement (at 2$\sigma$ level) with those obtained from the analysis of \ClI. For each component we tied the Doppler parameters of metal species using microturbulence assumption as $b^2 = b^2_{\rm turb} + 2k_BT/M$, where $M$ is the atomic mass of the species and $k_B$ is the Boltzmann constant. The turbulence Doppler parameter, $b_{\rm turb}$, and kinetic temperature of the gas, $T$, were left free during the fitting process. Our best fit is overlaid on the data in Fig.~\ref{metals}, and the corresponding parameters are given in Table~\ref{fit_res}. We note that a fit with independent $b$ parameters for each metal species gives column densities that are in agreement within 1$\sigma$ with the 'tied' fit, but have larger individual uncertainties.  We provide here below some comments about individual species.
  
\subsubsection{Zn\,{\sc ii}}
\label{Zn}
We estimate the column density of \ion{Zn}{ii} from the $\ion{Zn}{ii}\,\lambda$2026 and 2062 lines. $\ion{Zn}{ii}\,\lambda$2026 is redshifted in a spectral region crowded with O$_2$ sky lines. Since the different exposures with $437+760$ settings (that cover this spectral region) were taken at different times, the velocity of the observatory with respect to the line of sight is different, while O$_2$ lines remain
stationary. As a result, we find that in three out of five exposures, the $\ion{Zn}{ii}\,\lambda$2026 line is affected by sky lines but not in the remaining two. We therefore used the combination of these two exposures to constrain the fit of \ion{Zn}{ii}. Due to small velocity extent of the metal profile ($\sim 50$ km/s), there is no contamination of $\ion{Zn}{ii}\,\lambda$2026 and 2062 by $\ion{Mg}{i}\,\lambda$2026 and $\ion{Cr}{ii}\,\lambda$2062, respectively. 

\subsubsection{Si\,{\sc ii}}
\label{Si}
To determine the column densities of \ion{Si}{ii} in the three main components we use the relatively
weak $\ion{Si}{ii}\,\lambda1808$ line. Other \ion{Si}{ii} lines, $\ion{Si}{ii}\,\lambda$1304, 1526 and 1260
lines are highly saturated and show at least four additional components. Using these lines (together with $\ion{Si}{ii}\,\lambda$1808 line) we estimate \ion{Si}{ii} column density in these additional components to be $\log N(\ion{Si}{ii})\sim15$. This is significantly lower than what is measured in the main components $\log N (\ion{Si}{ii}) \sim15.75$. The situation is similar for other species such as \ion{Fe}{ii}. 

\subsubsection{S\,{\sc ii}}
\label{S}

\begin{figure*}
	\includegraphics[clip=,width=1.0\hsize]{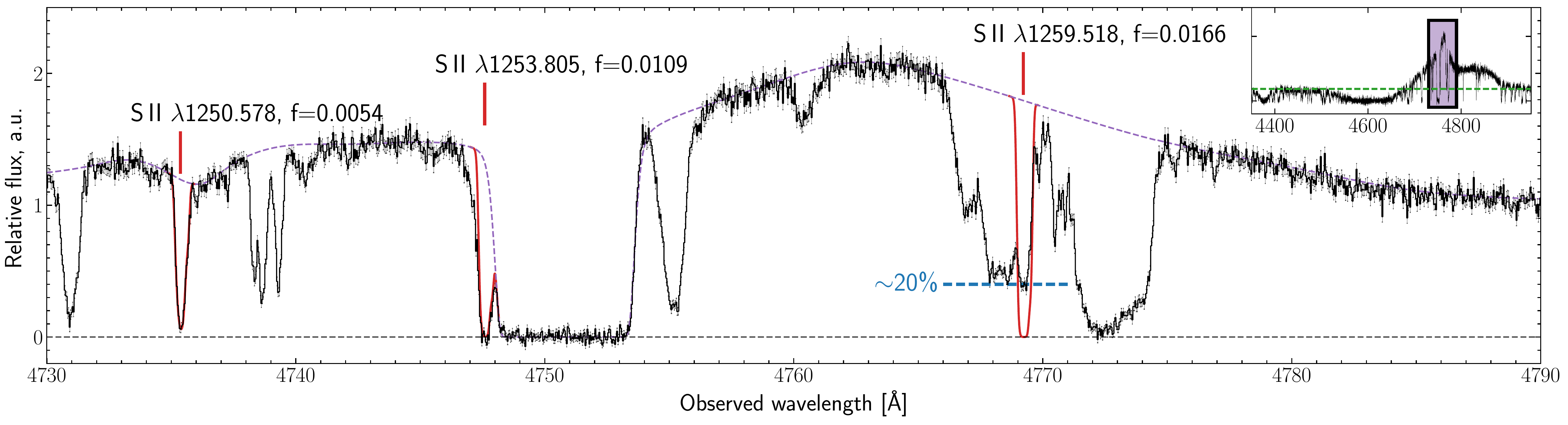}
	\caption{Portion of the UVES spectrum around the positions of S\,{\sc ii} absorption lines. The red lines show the calculated S\,{\sc ii} profiles based on the best-fitting parameters of $\ion{S}{ii}\,\lambda$1250. The dashed violet line indicates the local QSO continuum used for \ion{S}{II} line fitting. The horizontal blue dashed line indicates the observed bottom of the S\,{\sc ii}$\lambda$1259 line, which is located at $\sim$20\% of the total flux in that region.  
	{\sl Inner panel:} portion of the UVES spectrum around Ly$\alpha$ and \ion{N}{V} emission lines. Green dashed line indicates the estimate of the intrinsic QSO continuum. The violet box defines the region shown in the outer panel.}
	\label{SII_partcov}
\end{figure*}

We estimate the column density of \ion{S}{ii} using $\ion{S}{ii}\,\lambda$1250 while two other transitions
($\ion{S}{ii}\,\lambda$1253 and $\ion{S}{ii}\,\lambda$1259) are partially blended with other lines. Interestingly, using the best-fitting parameters obtained from nearly saturated $\ion{S}{ii}\,\lambda$1250 we found that the calculated profile of $\ion{S}{ii}\,\lambda$1259, whose oscillator strength is three times larger than that of $\ion{S}{ii}\,\lambda$1250, is deeper than the observed one. This is very likely the consequence of a partial coverage of the Ly-$\alpha$ emission line region of the quasar by the \ion{S}{ii}
absorbing region, since $\ion{S}{ii}\,\lambda$1259 is redshifted on the top of the quasar Ly-$\alpha$ emission line (see Fig.~\ref{SII_partcov}). Partial coverage has already been detected in at least two H$_2$-bearing intervening systems \citep[e.g.][]{Balashev2011, Klimenko2015}. From the residual flux at the bottom of the $\ion{S}{ii}\,\lambda$1259 line, we estimate that about $\sim 20$ per cent of the quasar photons
at the corresponding wavelengths are leaking through the S\,{\sc ii} absorbing region of the DLA. Since Ly-$\alpha$ emission contributes to $\sim 80$ per cent of emitted photons at that wavelength\footnote{The rest arising from the compact continuum emission close to the central black hole (hence very likely completely covered as also indicated by the saturated $\ion{S}{ii}\,\lambda$1253 line, see Fig.~\ref{SII_partcov}).}, this means that about $25$ per cent of the Ly-$\alpha$ photons arise from regions not covered by the DLA. Similar partial coverage (by H\,{\sc i}) of Ly-$\alpha$ quasar emission has been found for proximate DLAs at the quasar redshift \citep[see][]{Finley2013}, where further investigation show that the Ly-$\alpha$ emission can be very spatially extended \citep{Fathivavsari2016, Borisova2016}. Therefore such situation can be quite common. 

\subsubsection{P\,{\sc ii}}
\label{PII}
Phosphorus is a particularly useful element for metal enrichment studies, since it is believed to be little dust depleted \citep{DeCia2016}. Unfortunately, the strongest absorption line of singly ionized phosphorus, $\ion{P}{ii}\,\lambda$1152 is affected by blends. $\ion{P}{ii}\,\lambda$1301 is partially blended and not saturated. Using a three-component fit with fixed redshift and Doppler parameter deduced from the overall metal fit, we obtained an upper limit on the total phosphorus column density of $\log N$(\ion{P}{ii})~$<13.72$. The corresponding upper limit on the metallicity $\mbox{[P/H]}<-1.28$ is consistent with low depletion of the phosphorus in the DLA at z=2.7865.

\subsubsection{Excited \ion{C}{II}* and \ion{Si}{II}*}
\label{Si_exc}

\begin{figure}
	\begin{center}
		\includegraphics[clip=,width=0.95\hsize]{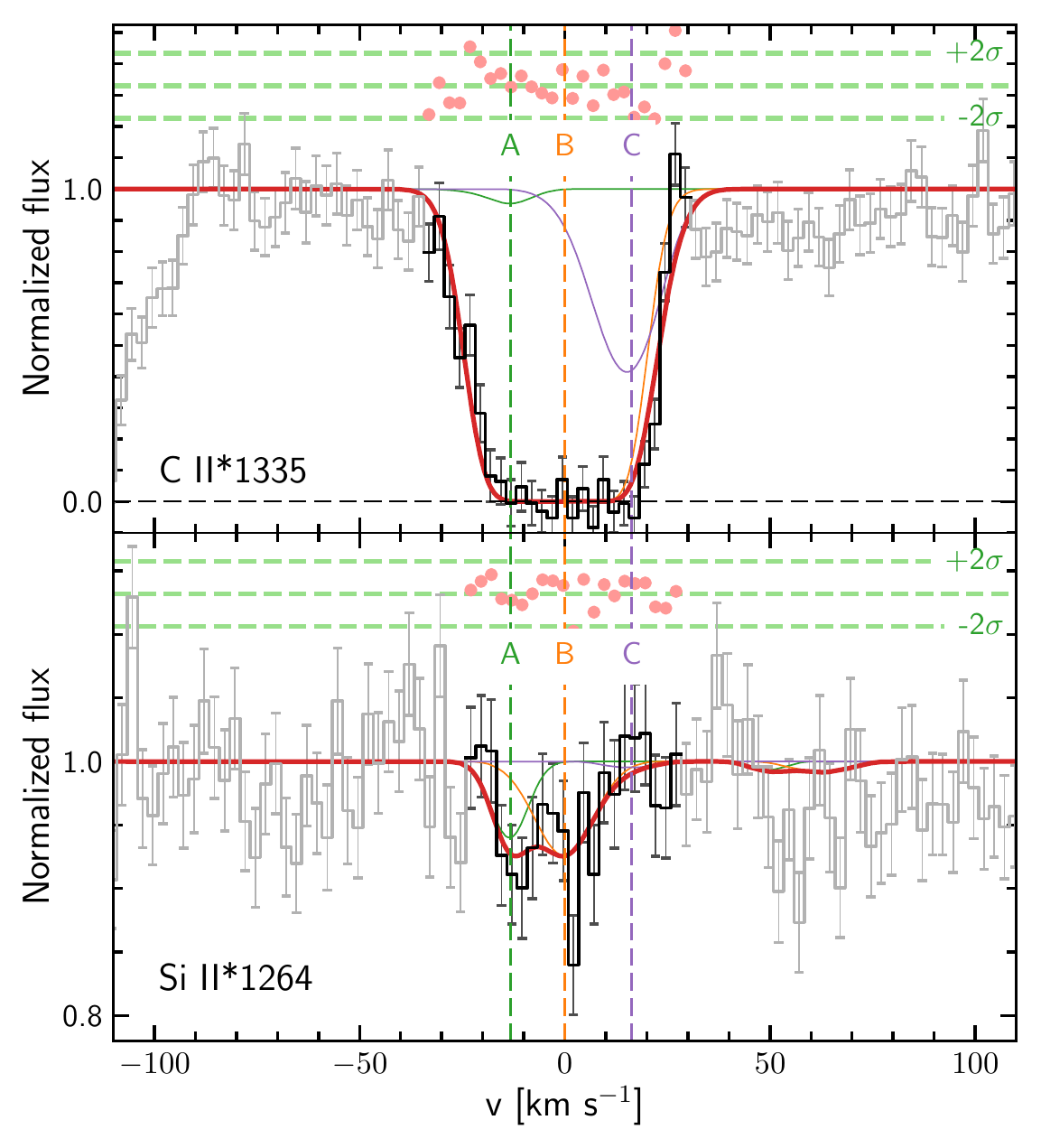}
		\caption{Fit to \ion{C}{II*} and \ion{Si}{II*} lines, associated with DLAs at $z=2.7865$.}
		\label{SiII_exc}
	\end{center}
\end{figure}

The detection of excited \ion{Si}{ii*} was recently reported in the two extremely strong DLAs (ESDLAs) at $z=2.21$ towards J\,1135$-$0010 \citep{Kulkarni2012} and at $z=2.34$ towards J\,2140$-$0321 \citep{Noterdaeme2015}. { \citet{Neeleman2015} detected \ion{Si}{ii*} in another seven DLAs, which also correspond to the high end of the \HI\ column density distribution}. The ratio $N(\ion{Si}{ii}*)/N(\ion{Si}{ii})$ can be used to estimate the number density (or electron density)
since the excited state of \ion{Si}{ii} is predominantly populated by collisions and depopulated by spontaneous transitions \citep{Srianand2000a}. Since the DLA studied here also has a very high column density ($\log N$(\HI)~$\approx 21.8$), we searched for \ion{Si}{ii*} absorptions lines, which we tentatively detect in the strongest transition \ion{Si}{ii}* $\lambda1264$ (see Fig.~\ref{SiII_exc}). Fixing the redshifts
and Doppler parameters to that obtained for \ion{Si}{ii}, our fitting procedure indicates $\log N(\ion{Si}{ii}*)=11.32^{+0.14}_{-0.21}$ and $\log N(\ion{Si}{ii}*)=11.62^{+0.08}_{-0.14}$ for components $A$ and $B$, respectively. Because the detection is only tentative, we rather consider the 2$\sigma$ upper limit on the total \ion{Si}{ii*} column density ($\log N(\ion{Si}{ii}*)<12$) in the following. This translates to
$N(\ion{Si}{ii}*)/N(\ion{Si}{ii}) \le 1.7\times10^{-4}$, i.e., similar to what is found towards J\,2140$-$0321 but an order of magnitude lower than the ratio measured towards J\,1135$-$0010.

We derive the column density of excited singly ionized carbon using the $\ion{C}{ii*}\,\lambda1335$ line
(see Fig.~\ref{SiII_exc}). Because this line is saturated, we used a three component fit with redshifts and
Doppler parameters fixed to the values derived from the joint fit of other metal lines. We found that most
of the \ion{C}{ii*} column density arises from component $B$ (see Table~\ref{fit_res}).

\subsection{Extinction of the background quasar's light}
\label{extinction}
\begin{figure*}
	\begin{center}
		\includegraphics[clip=,width=0.95\hsize]{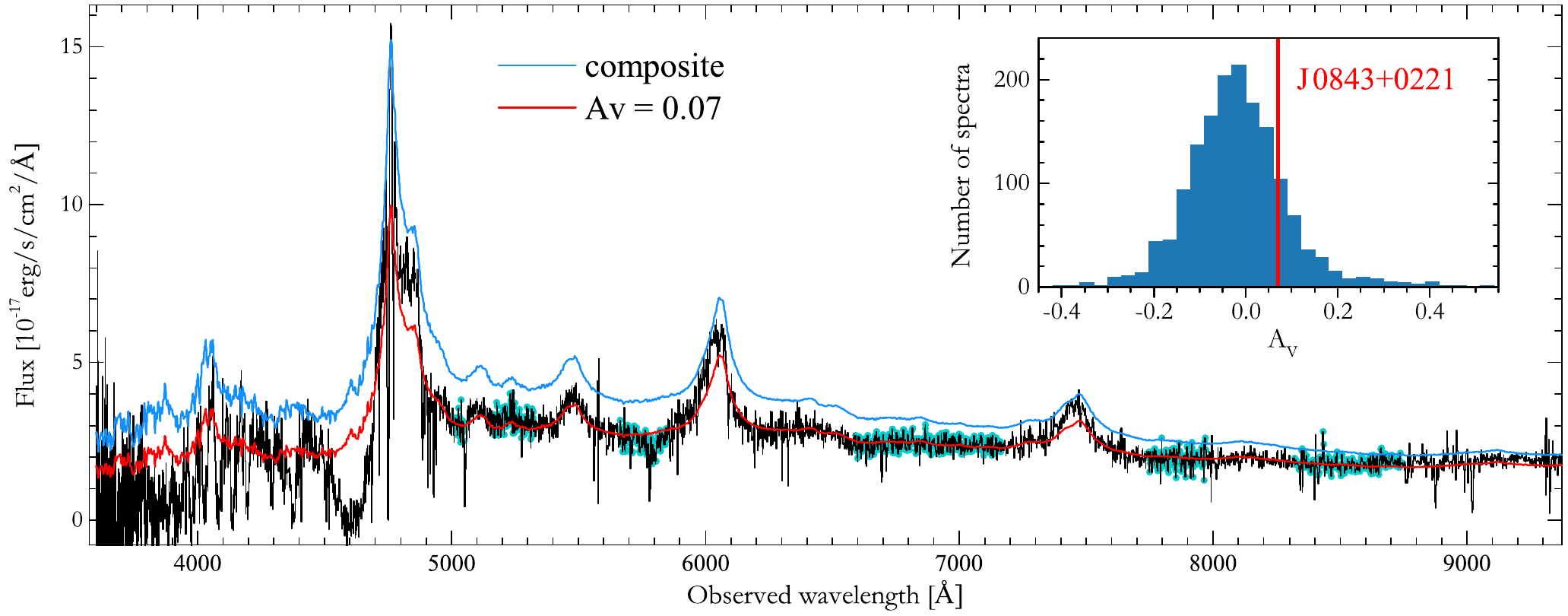}
		\caption{SDSS spectrum of \qso (black) compared to the unreddened SDSS composite (blue) and the composite spectrum reddened by our best-fitting value $A_V=0.07$. The cyan points indicate regions of the spectrum that were used to constrain the continuum fit. The inlay shows the distribution of measured $A_V$ for the control samples of spectra without having DLAs. This provides the systematic error of $A_V$ measurement. The red line indicates the extinction measured in the DLA at z=2.7865 towards \qso. In spite of the large \HI\ and H$_2$ column densities the measured upper limit A$_V < 0.1$ is a factor of $\sim30$ lower than the value expected from the extrapolation of relation between $A_V$ and hydrogen column density obtained in the SMC.}
		\label{qso_ext}
	\end{center}
\end{figure*}

We used the flux-calibrated SDSS spectrum of \qso\ to measure the extinction due to the H$_2$-bearing DLA at $z=2.7865$. We followed a standard procedure similar to that presented in \citet{Srianand2008}: we fitted the observed spectrum of \qso\ using the SDSS composite spectrum from \citet{VandenBerk2001}, redhifted to $z_{\rm em}=2.92$ and reddened using an extinction law applied at $z_{\rm abs}=2.7865$. Since no 2175~{\AA} bump is detected, we used the average Small Magellanic Cloud (SMC) extinction law \citep{Gordon2003} and found a best-fitting value of $A_V=0.07$ (Fig.~\ref{qso_ext}). In order to estimate the systematic uncertainty due to intrinsic variations of the spectral energy distribution of quasars, we used a control sample of quasar spectra from SDSS DR9 with no detected DLA and with emission redshifts in the range
$\Delta z= 0.05$ around that of \qso. We applied the same fitting procedure to each spectrum in the control sample and obtained the distribution of measured $A_V$ shown in the inlay of Fig.~\ref{qso_ext}. The median and dispersion in the control sample are $-0.02$ and $0.10$, respectively, which give the zero-point and systematic error for the measured $A_V$ in \qso. We thus obtain $A_V = 0.09\pm0.10$ towards \qso.

Such a low amount of extinction is incompatible with the value expected from extrapolating the canonical extinction-to-gas ratio seen in the MW: for the total hydrogen column density $\log N({\rm H}) = \log N(\mbox{\HI}) + \log 2N(\mbox{H}_2))= 21.99^{+0.08}_{-0.07}$, we would expect $A_V \approx 5$ \citep{Watson2011}. Using the SMC and Large Magellanic Cloud (LMC) extinction-to-gas ratio \citep{DeCia2013} does not change significantly the expected value. The low $A_V$ measured here is however not surprising given the low metallicity of the DLA. Using the observed metallicity and depletion factor, we estimate the dust-to-gas ratio to be $\kappa = \rm [Zn/H]\left(1-10^{[Fe/Zn]}\right) \approx 0.03$, i.e. the expected $A_V$ should be reduced by a factor of about 30 compared to the above extrapolations. We then expect $A_V = 5\times 0.03 \approx 0.15$, i.e. very close to the observed value. 

\section{Physical conditions}
\label{conditions}

\subsection{Metallicity and depletion pattern}
\label{metal}
For each element we determine the abundance relative to solar values from \citet{Asplund2009}, without applying any ionization correction (i.e. we assume all metals to be in their main ionization state). This is a fairly reasonable assumption given the large \HI\ column density in this DLA. Since we cannot determine the hydrogen column density in individual components, we derive gas-phase abundances using the total hydrogen column density $N({\rm H})=21.99^{+0.08}_{-0.07}$ and the total metal column densities, see Table~\ref{t:abund}.

\begin{table}
  \centering
  \addtolength{\tabcolsep}{-3pt}
  \setlength\extrarowheight{3pt}
  \caption{Summary of the overall gas-phase abundances \label{t:abund}}
  \begin{tabular}{l c}
    \hline \hline
    Species ($X$) & [$X$/H] \\
    \hline
    \ion{Cl}{i}  & $-1.86^{+0.21}_{-0.10}$ \\
    \ion{Mg}{ii} & $-1.65^{+0.09}_{-0.10}$ \\
    \ion{Si}{ii} & $-1.73^{+0.08}_{-0.10}$ \\
    \ion{S}{ii}  & $-1.54^{+0.08}_{-0.08}$ \\
    \ion{Ti}{ii} & $-1.96^{+0.10}_{-0.11}$ \\
    \ion{Cr}{ii} & $-2.40^{+0.09}_{-0.10}$ \\
    \ion{Mn}{ii} & $-1.75^{+0.10}_{-0.10}$ \\
    \ion{Fe}{ii} & $-2.53^{+0.08}_{-0.10}$ \\
    \ion{Ni}{ii} & $-2.52^{+0.09}_{-0.10}$ \\
    \ion{Zn}{ii} & $-1.52^{+0.08}_{-0.10}$ \\
    \hline
  \end{tabular}
  \addtolength{\tabcolsep}{3pt}
 \end{table}

The observed gas-phase abundances of zinc and sulphur are usually considered as good indicators of the ISM metallicity since these are non-refractory elements for which dust depletion corrections are expected to be negligible. Their abundances, $\rm [Zn/H]=-1.52^{+0.08}_{-0.10}$ and $\rm [S/H]=-1.54^{+0.08}_{-0.08}$, are indeed in very good agreement. 

\begin{figure}
	\begin{center}
		\includegraphics[clip=,width=0.95\hsize]{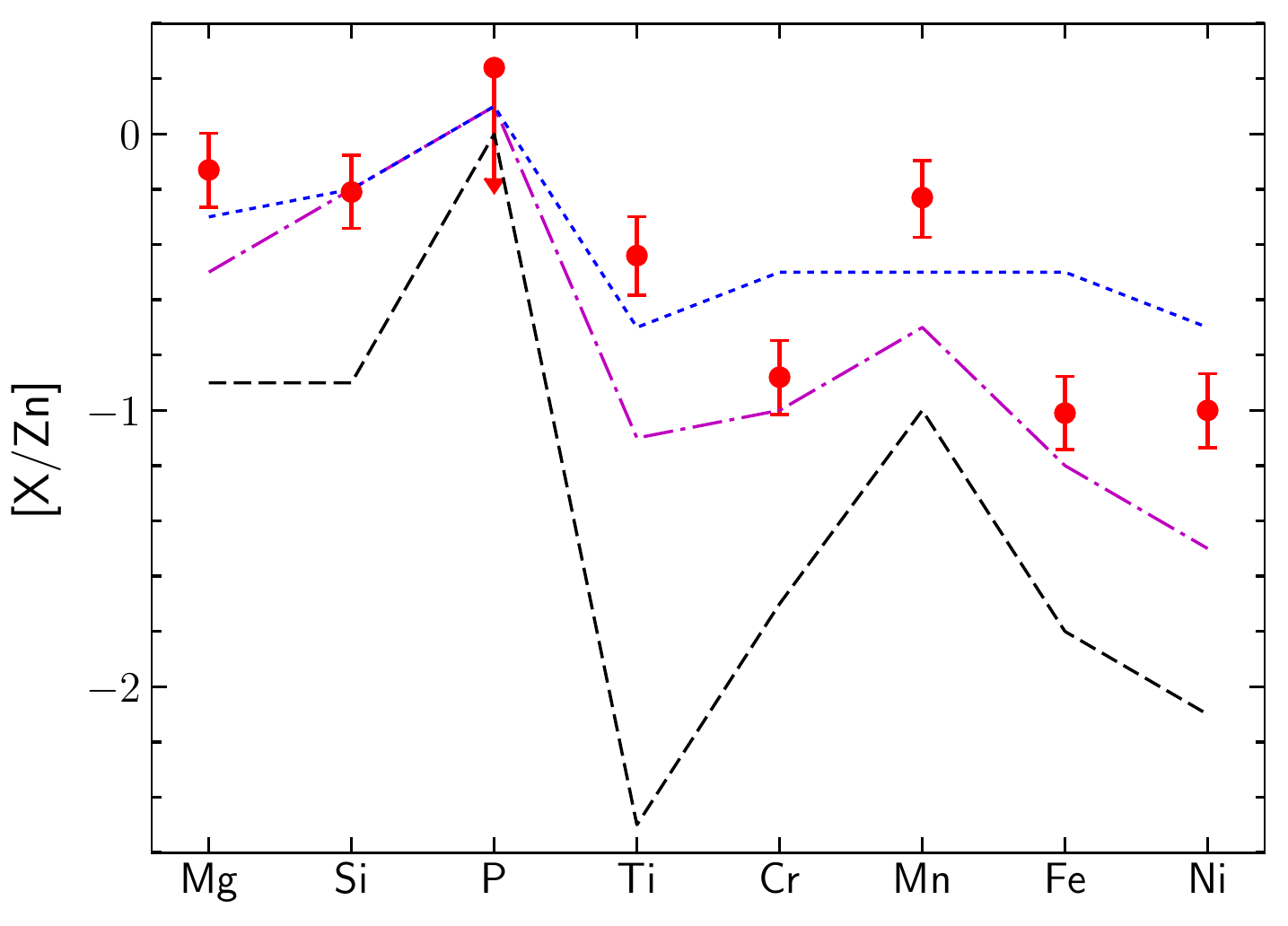}
		\caption{Depletion pattern of different elements in the DLA at z=2.7865 (red dots with error bars).
			Blue dotted, magenta dash-dotted and black dashed lines show the typical depletion pattern inferred for MW halo, warm and cold ISM, respectively \citep{Welty1999}. { The depletion pattern is observed to be mild compared to that of the cold ISM gas in the Galaxy.}
				}
		\label{depletion}
	\end{center}
\end{figure}

Other elements have lower observed gas-phase abundances. In particular, \ion{Fe}{ii}, \ion{Cr}{ii}, \ion{Ni}{ii} and \ion{Ti}{ii} are refractory elements that are known to severely deplete on to dust grains. In Fig.~\ref{depletion}, we compare the observed depletion pattern in our DLA with typical values seen in the MW halo, the warm and the cold ISM \citep{Welty1999}. The observed depletion pattern is found to be intermediate between those observed for the warm and halo MW gas and to be much less depleted than the cold ISM gas in the MW. As noted by \citet{Noterdaeme2007}, this is most likely a metallicity effect and does not mean that the DLA gas is either warm or in a halo. Indeed, the measured \ion{Fe}{ii} depletion, $\rm [Fe/Zn] = -1.01\pm0.13$ is larger than typical values of $\sim-0.25$ observed in DLAs at similar metallicities
in the overall population of high-$z$ DLAs \citep{Rafelski2012, Quiret2016, DeCia2016} or in the sub-population of ESDLAs \citep{Noterdaeme2014}. Similarly, high depletion level, $\rm [Fe/P]=-1.21^{+0.16}_{-0.17}$, (for such a low-metallicity system) has recently been observed in the ESDLA ($\log N(\HI) = 22.4\pm0.1$) towards J\,2140$-$0321 \citep{Noterdaeme2015}. The two systems actually share similar properties: the DLA towards J\,2140$-$0321 also has low metallicity ([P/H]$\sim -1.05$) and high H$_2$ column density ($\log N({\rm H}_2) = 20.13\pm0.07$). 

\begin{figure}
	\begin{center}
		\includegraphics[clip=,width=0.95\hsize]{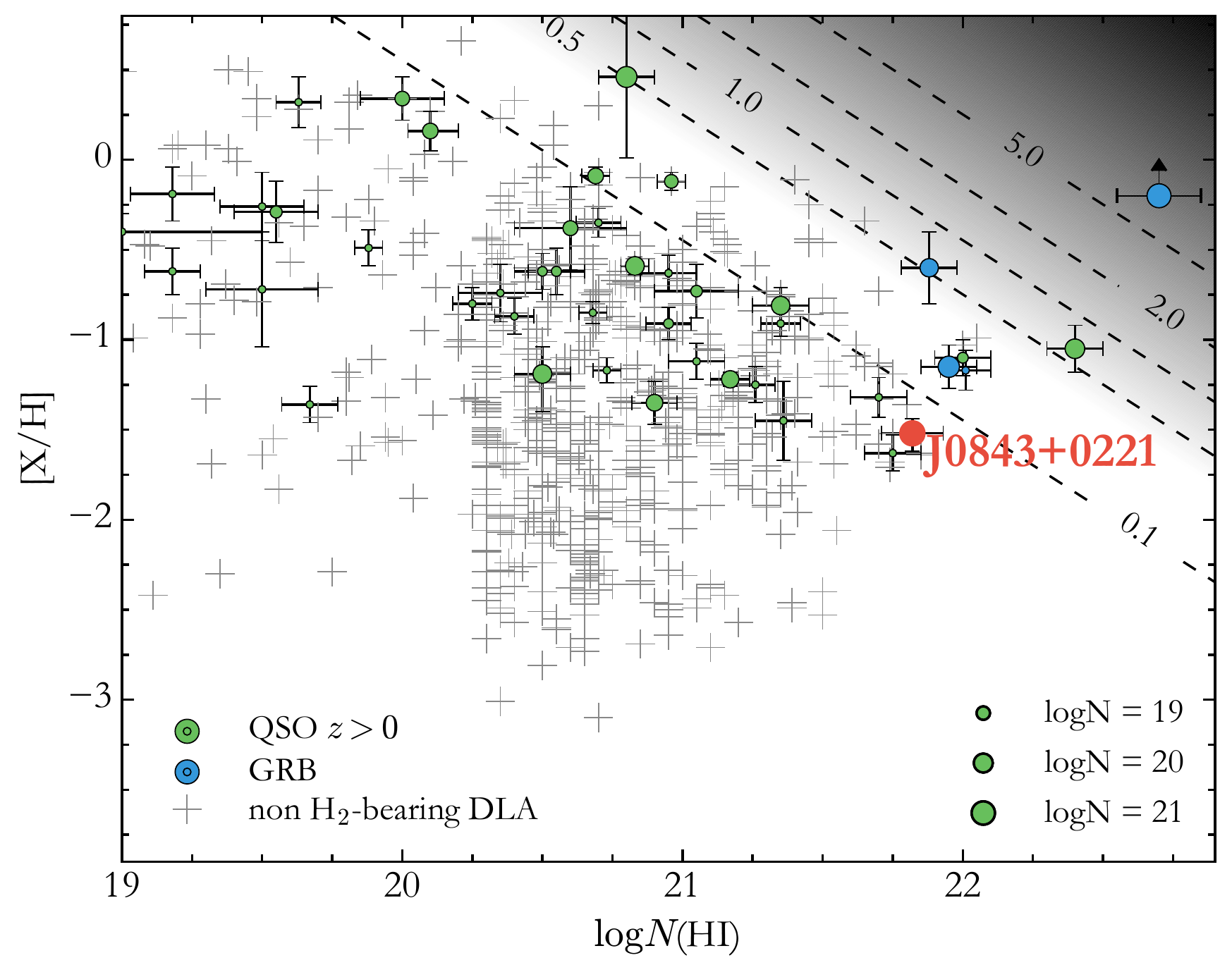}
		\caption{Metallicity versus \HI\ column density in DLAs. The green and blue circles indicate the H$_2$-bearing DLAs at high redshifts detected towards QSO and in GRB afterglows, respectively. The size of each circle depends on H$_2$ column density. The grey crosses correspond to non-H$_2$-bearing DLAs. The dashed black lines show constant value of $A_V$ calculated in the assumption to fixed scalefactor between metallicity and dust abundance corresponds to the MW measurements. The shaded region reflects systematic bias due to obscuration.}
		\label{MevsHI}
	\end{center}
\end{figure}

It is evident that the dust content scales not only with depletion, but also with the gas-phase metallicity. Fig.~\ref{MevsHI} shows the location of H$_2$-bearing DLAs in the metallicity$-N(\HI)$ plane. There are 40 published $z>0$ quasar H$_2$ absorption systems
\citep{
	Levshakov1985,  
	Ge1997,         
	Ge2001,         
	Ledoux2002,     
	Levshakov2002,  
	Reimers2003,    
	Ledoux2003,     
	Cui2005,        
	Ledoux2006,     
	Petitjean2006,  
	Noterdaeme2007, 
	Noterdaeme2008, 
	Srianand2008,   
	Jorgenson2009,  
	Malec2009,      
	Noterdaeme2010, 
	Jorgenson2010,  
	Srianand2010,   
	Fynbo2011,      
	Guimaraes2012,  
	Srianand2012,   
	Crighton2013,   
	Oliveira2014,   
	Balashev2015,   
	Noterdaeme2015, 
	Muzahid2015,    
	Muzahid2016,    
	Krogager2016,   
	Noterdaeme2017} 
and this paper. Also there are four H$_2$-bearing systems detected in GRB afterglows \citep{Prochaska2009b, Kruhler2013, Friis2015, DElia2014}. As observed by several authors \citep[e.g.][]{Petitjean2006}, all H$_2$ detections are located in the upper part of the diagram, above a  threshold value for overall metallicity
$\sim$0.02~$Z_{\rm \odot}$ (i.e H$_2$-bearing DLAs have statistically higher metallicities than overall population of DLAs). Among H$_2$-bearing DLAs, the system towards \qso\ has one of the lowest metallicity in spite of having the largest H$_2$ column density. This could be due to a dust selection effect against DLAs that would have similar hydrogen column densities but higher metallicities. Indeed, in our Galaxy, absorption systems with $\log N({\rm H}) \sim 10^{22}$ cm$^{-2}$ are expected to have $A_V \gtrsim3$. Such an extinction would definitely preclude such systems from the SDSS. There are only two exceptional cases: the measurement in the DLA associated with GRB~080607 gives $A_V\approx3.2$ mag \citep{Prochaska2009b} and the $z=0.89$ lensing system towards PKS\,1830$-$211 where very high column density of H$_2$ is expected given the plethora of other detected molecules \citep{Muller2011, Muller2014}. In this latter peculiar case, the background blazar is completely extinguished in the optical bands and was selected from its millimetre emission and the foreground lensing galaxy \citep[see][]{Wiklind1996}. 

As mentioned previously, chlorine is tightly connected with H$_2$, with all chlorine is neutral in the H$_2$-bearing gas while is singly ionized elsewhere. In addition, we expect little depletion of chlorine \citep{Neufeld2009} for the low metallicity and reddening (see Sect.~\ref{extinction}) measured in this system. We can therefore use the abundance of neutral chlorine to constrain the metallicity in the H$_2$-bearing gas assuming its molecular fraction, or, conversely, make assumptions on the metallicity in that phase to constrain the molecular fraction \citep{Balashev2015}. The measured $\log N(\ClI) =13.63^{+0.20}_{-0.05}$ translated to a molecular fraction in the H$_2$ phase of 0.8, under the assumption that the metallicity in that phase equals the overall metallicity. It is also likely that the metallicity is actually higher in the H$_2$-bearing gas, meaning that the derived molecular fraction can be considered as
a lower limit, i.e. the cloud is almost fully molecularized. This is what is expected from the modelling  with high H$_2$ column density clouds (but see \citealt{Noterdaeme2017} where a high cosmic ray ionization rate can still maintain a non-unity molecular fraction even after the H\,{\sc i}-to-H$_2$ transition).

\subsection{Kinetic temperature}
\label{T01}

\begin{figure}
	\begin{center}
		\includegraphics[clip=,width=0.95\columnwidth]{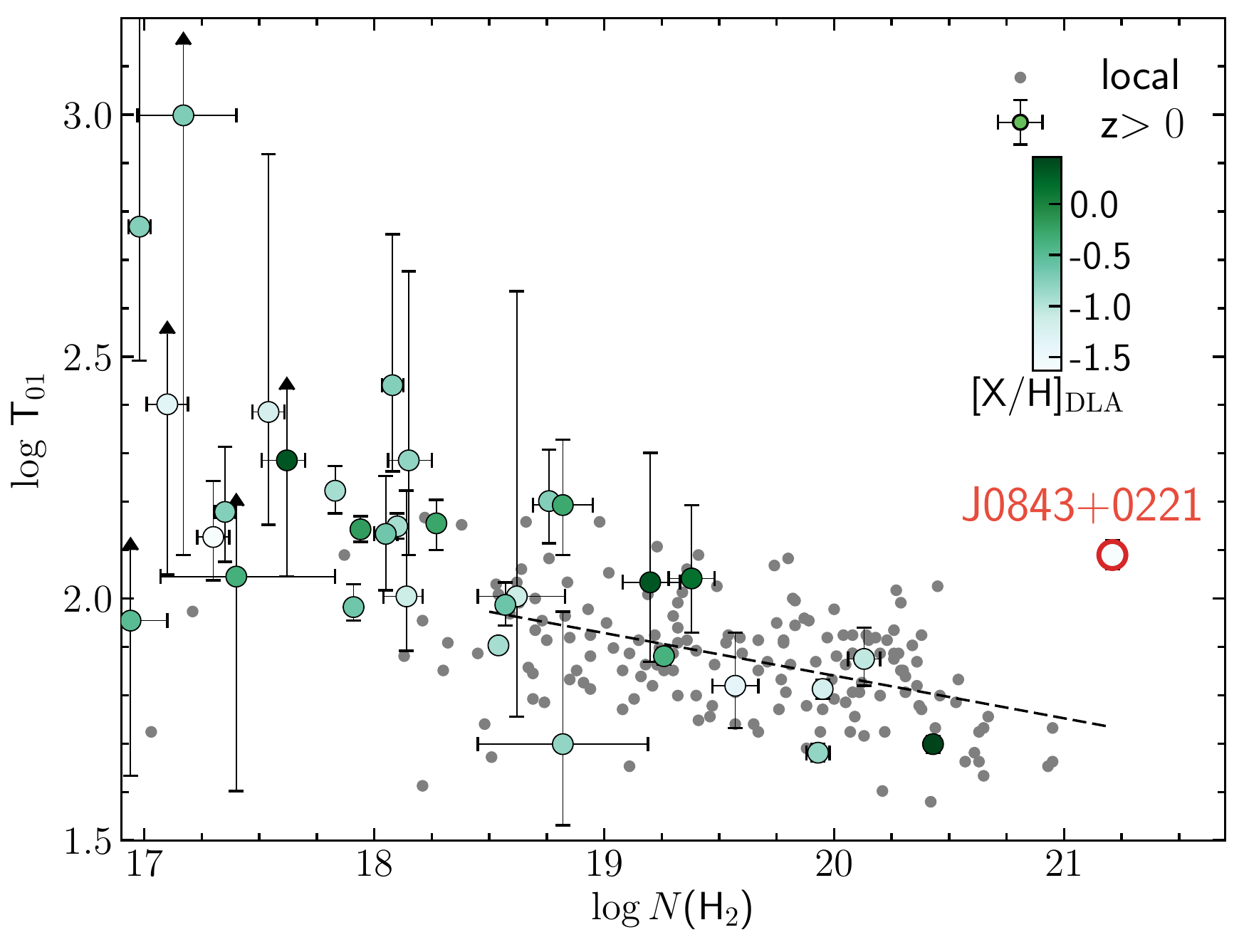}
		\caption{Measurements of the H$_2$ excitation temperature, $T_{01}$ at different $N(\rm H_2)$.
                  The grey points represent measurements in the Local Group (MW, SMC and LMC). 
                  Filled circles represent measurements at $z>0$, where the colour corresponds to the average metallicity of each associated DLA. The measurement in the DLA towards \qso\ is at least a factor of 2 higher than the value expected from extrapolating the trend of decreasing $T_{01}$ with increasing $N(\rm H_2)$.
              	}
		\label{H2Tkin}
	\end{center}
\end{figure}

In Sect.~\ref{H2}, we found the excitation temperature of the first excited level of H$_2$ to be $T_{01} = 123^{+9}_{-8}$\,K. This is known to provide a good estimate for the kinetic temperature of the gas \citep[e.g.][]{Roy2006}. In Fig.~\ref{H2Tkin}, we compare this value with measurements of $T_{01}$ in both distant DLAs and in the local Universe. The data for the local Universe were taken from \cite{Savage1977} for the MW, \cite{Gillmon2006} for high-latitude MW clouds and \cite{Welty2012} for the SMC and LMC. The excitation temperature in the DLA at z=2.7865 is similar to what is typically measured in high-$z$ and low-$z$ H$_2$-bearing DLAs ($\langle T_{01} \rangle\sim 150$ K; \cite{Srianand2005,Muzahid2015}) but
higher than what is measured locally. However, as can be seen in Fig.~\ref{H2Tkin}, $T_{01}$ clearly decreases as a function of $N$(H$_2$) so that the difference in the 'average' temperatures between the local and distant measurements is mostly due to a difference in the probed H$_2$ column densities. In the overlapping range of column densities (e.g. around $\log N ({\rm H}_2)\sim 18.5-20.5$), temperature measurements agree well between $z=0$ and $z>0$.  

It is striking that the $T_{01}$ value measured in the DLA towards \qso\ is significantly higher than $T_{01}$ in other H$_2$ strong systems. For $\log N({\rm H}_2)>21$, we would expect $T_{01} \lesssim 50$\,K when we observe more than twice this value. A possible explanation of this significant difference resides in the low metallicity of this system. The cooling through the metal line emission (mostly [C\,{\sc ii}]157$\mu m$) can be much less efficient than at high metallicity and the equilibrium is shifted to higher temperatures. This possibility is supported by recent calculations of the metallicity dependence of temperature states in the neutral medium \citep[e.g.][]{Glover2013}.  However, there is no apparent trend in the measured $T_{01}$ as a function of metallicity (see colour code of the circles in Fig.~\ref{H2Tkin}). Since the system we are studying is unique both in terms of metallicity and $N$(H$_2$), it would be highly desirable to obtain additional measurements of molecular gas temperature in low-metallicity environments.

\subsection{HD/2H$_2$ ratio}
The relative abundance of HD to H$_2$ molecules is known to be sensitive to the local conditions in the gas \citep{Balashev2010,Liszt2015}. Because of the low metallicity of our system, we can expect the astration of deuterium to be negligible \citep[see][]{Dvorkin2016}. The intrinsic deuterium abundance should therefore be close to the primordial value, the latter being well constrained from the observations of atomic D\,{\sc i} and H\,{\sc i} lines in Lyman limit systems and (sub)-DLAs (e.g. \citealt{Pettini2012, Noterdaeme2012b, Cooke2016, Balashev2016}) or inferred from the analysis of the primordial anisotropy of CMB using the results of the big bang nucleosynthesis \citep{Ade2014}.  We measure $N$(HD)/2$N$(H$_2$)$=6.9^{+2.9}_{-3.8}\times10^{-5}$. This value is consistent within errors with the measurements in Q\,1232+082 and Q\,0812+3208 and the primordial D/H ratio. Such a high HD/2H$_2$ ratio is expected to occur when all the D and H are in their molecular form. The abundance of neutral chlorine also indicates a very high molecular fraction (close to unity) in the H$_2$-bearing gas. However, since the measured HD/2H$_2$ ratio is marginally higher than the intrinsic isotopic ratio (D/H), one can assume that the molecular fraction of hydrogen is slightly below unity.

\subsection{Number density}
\label{numbdens}
The relative populations of the fine-structure levels of \ion{C}{i}, \ion{C}{ii}, \ion{Si}{ii} and the rotational levels of HD can be used to estimate the number density in the medium. For the typical conditions of the cold neutral medium, these excited levels are mostly populated by collision with molecular and atomic hydrogen (depending on the actual molecular fraction in the medium) and depopulated by collisions and spontaneous transitions. In most cases, collisions with electrons can be neglected. For example, if the \ion{Si}{ii}* is mostly populated by electron collisions, then the measured \ion{Si}{ii}*/\ion{Si}{ii} ratio would imply an ionization fraction $>0.01$ which is very unlikely given the cold and molecular nature of the cloud. Similarly, radiative pumping by the UV background field is only important at low densities \citep{Silva2002}. A very high value of the UV field, more than 100 times the mean Galactic value \citep{Draine1996} would be needed to explain the observed populations of the \ion{C}{i} fine-structure levels. Therefore we include neither the electron collision nor the radiative pumping by UV photons in the analysis. This allows us to reduce the number of parameters in the calculations. In turn, we do take into account direct excitation by the CMB radiation. This does not add any additional free parameter since the 
redshift of the system ($z \approx 2.7865$) is known and hence the value of the CMB field is fixed under standard cosmology. In any case, we found that radiative transitions by the CMB provide only a very small contribution to the excitation of the above species, even for \ion{C}{i}, whose transition energies between fine-structure levels are close to the CMB temperature at the DLA redshift.

To calculate how the relative populations of the fine-structure levels of \ion{C}{i}, \ion{C}{ii}, \ion{Si}{ii} and the rotational levels of HD depend on the physical conditions in the gas we used the code which is similar to the \textsc{PopRatio} \citep{Silva2001, Silva2002} and \textsc{RADEX} \citep{vanderTak2007} codes. We used data mostly provided along with \textsc{PopRatio}, which includes energy levels \citep{Moore1970}, transition probabilities \citep{Galavis1997} and collisional coefficients \citep{Schroder1991, Staemmler1991, Abrahamsson2007} for \ion{C}{i} and energy levels \citep{Moore1970}, transition probabilities \citep{Galavis1997} and collisional coefficients \citep{Wiesenfeld2014} for \ion{C}{ii} and \ion{Si}{ii}. The atomic data for HD were taken from \cite{Flower2000}. 

Neglecting collision with electrons, UV pumping and fixing the CMB field, we are left with three parameters
to calculate the relative populations of excited levels: kinetic temperature, number density and molecular fraction. We then perform calculations using two extreme cases for the local molecular fraction: $(i)$ $f=0.3$, which corresponds to the overall observed molecular fraction and is hence a strict lower limit since it assumes all \ion{H}{i} in the DLA to be associated with the H$_2$-bearing phase and $(ii)$ $f=1$, where the only collision partner is H$_2$. Note that the abundances of neutral chlorine and HD indicate that we should be closer to the fully molecularized case. We also assume the medium to be homogeneous, i.e. the calculation is performed using a one-zone model: the molecular fraction, number density and temperature are constants throughout the cloud.

\begin{figure}
	\centering
	\begin{tabular}{cc}
		\includegraphics[clip=,width=0.95\hsize]{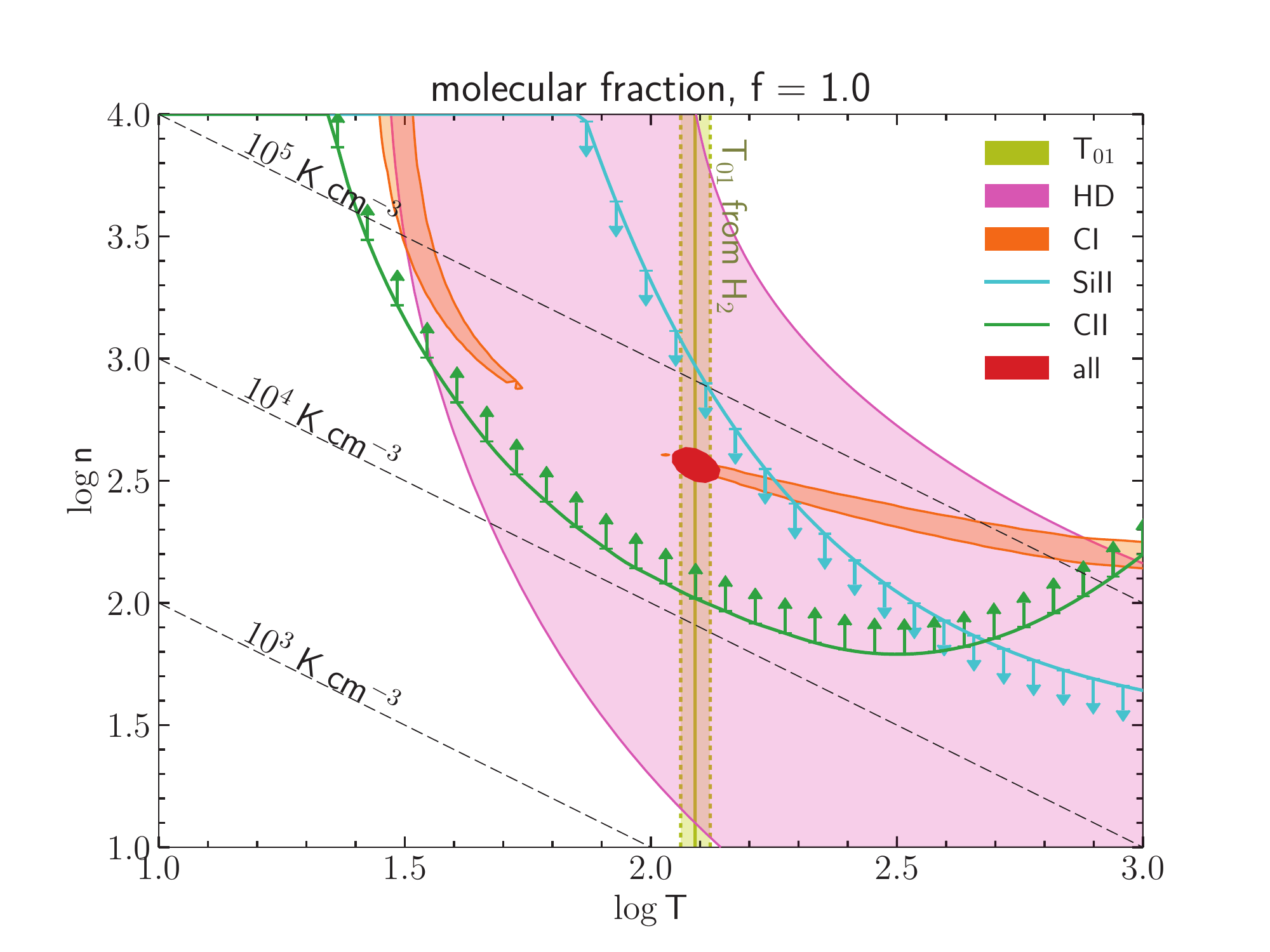} \\ 
		\includegraphics[clip=,width=0.95\hsize]{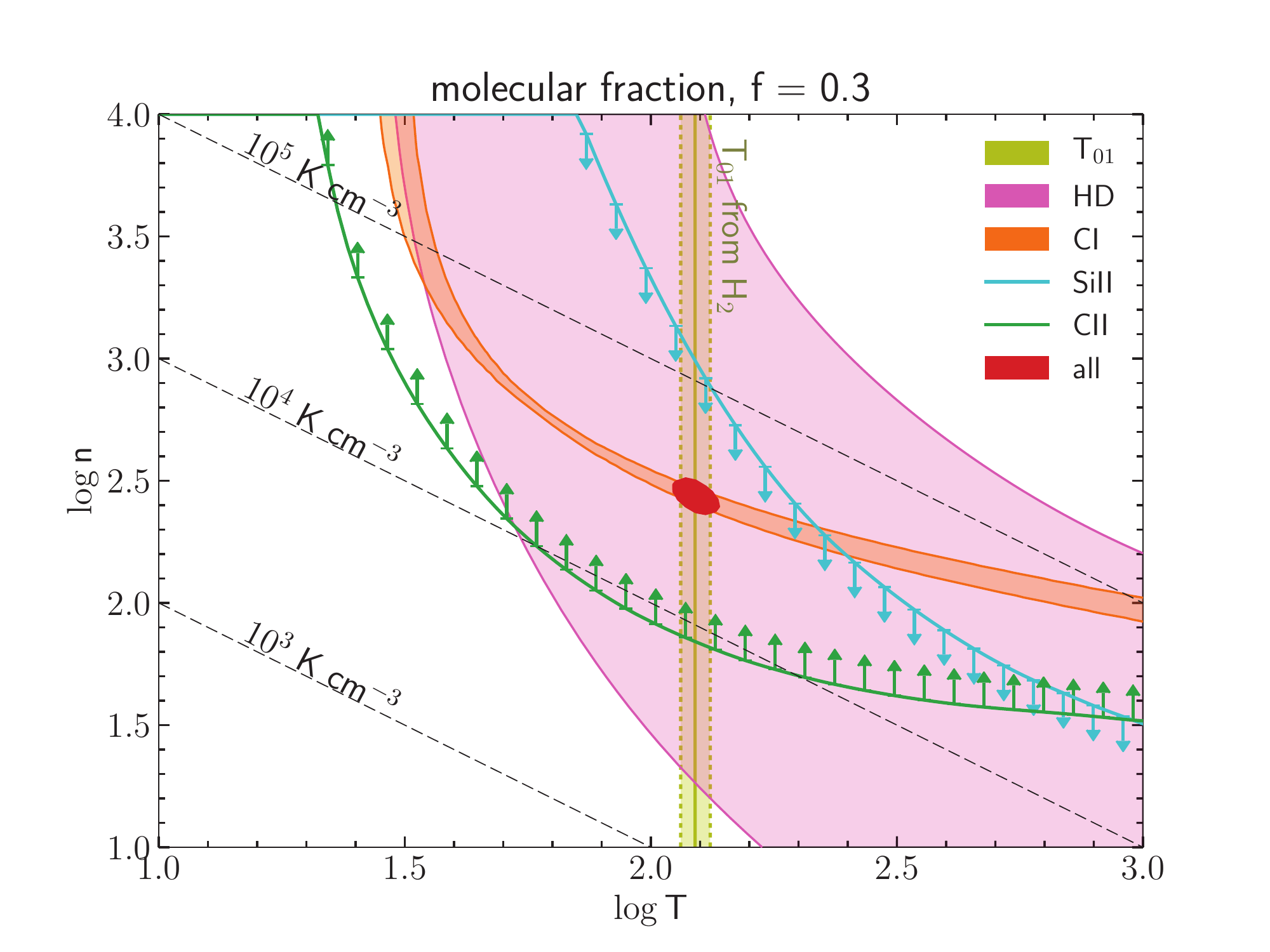} \\ 
	\end{tabular}
	\caption{
		Constraints on the number density and kinetic temperature obtained from different excited species. Constraints from \CI\ and HD are in orange and purple, respectively, while the excited ionized silicon and carbon provide respectively upper and lower limits on the density for a given temperature. The red region shows the joint estimate when also considering the excitation of H$_2$. The top and bottom panels are for molecular fraction $f$=1.0 and $f$=0.3, respectively. The estimated number density in the DLA at z=2.7865 is higher than typical values obtained in the diffuse ISM both in the local Universe and at high redshifts.
      		}
	\label{Tdens}
\end{figure}

To compare the calculated values with the observed populations we used a maximum likelihood function. Since for some species we do not have good observational constraints on the column densities for individual components, in the following we consider the total column densities instead. This should give a good approximation of average physical conditions in the gas. We note indeed that for \ion{C}{I}, the population ratios in both components -- on which we have good constraints -- are very similar.  While we have a measurement of \ion{C}{ii} in its excited state (i.e. \ion{C}{ii*}), we do not have a measurement of the ground state. From the zinc metallicity, and assuming solar intrinsic abundances, we estimate $\log N(\ion{C}{ii}) = 16.93$. We considered this value as an upper limit since carbon is likely more depleted
in the diffuse ISM than zinc. Since \ion{Si}{ii*} is only tentatively detected, we also treat \ion{Si}{ii*} column density as an upper limit. The results of calculations are shown in Fig.~\ref{Tdens}.

The allowed regions in the temperature$-$density plane for the different species overlap for the two assumed
molecular fractions, but in different regions of that plane. When adding the constraint on the kinetic temperature independently derived in Sect.~\ref{T01}, the number density is found to be $\log \rm n = 2.58^{+0.03}_{-0.06}$ and $\log \rm n = 2.42^{+0.06}_{-0.03}$ { (where $\rm n$ is in units of cm$^{-3}$)} for the $f=1$ and $f=0.3$ cases, respectively. This is significantly higher than what is generally estimated in the neutral medium $n \sim 0.1 - 10\, \mbox{cm}^{-3}$ and also higher than what is usually derived in other molecular-rich DLA systems \citep[e.g.][]{Noterdaeme2017}. 

\subsection{Thermal pressure}
Using the estimates of the number density and kinetic temperature of the cloud, we derive the thermal pressure, $P = n T = 4.7^{+0.5}_{-0.6} \times 10^4$~cm$^{-3}$\,K. While the thermal pressure is not necessarily dominating the total pressure (that includes radiation, turbulence and magnetic field pressures),
it is expected to correlate with other parameters of ISM, like turbulent pressure \citep{Joung2009} as well as
heating and star formation rates \citep{Ostriker2010}. 

In our Galaxy typical thermal pressures are found to be in the range from $2\times10^{3}$ to $10^4$ K\,cm$^{-3}$ \citep{Jenkins2011}. Similar values were recently obtained in a sample of nearby galaxies \citep{Herrera-Camus2017}. However, \citet{Welty2016} measured slightly higher values in the Magellanic Clouds. This is consistent with the theoretical predictions that higher pressures are required for stable cold, neutral clouds in environments with low metallicity, low dust-to-gas ratios and enhanced radiation fields. Higher pressures are therefore also expected in DLAs at high redshifts.  However, the measurements by \citet{Jorgenson2010}  using \ion{C}{I} and \citet{Neeleman2015} using \ion{C}{II} and \ion{Si}{II} give pressures consistent with MW values. We note that the measurements by \citet{Neeleman2015} are consistent with an increase of thermal pressure with increasing $N (\rm H)$, despite a large uncertainty on individual measurements.

\begin{figure}
	\begin{center}
		\includegraphics[clip=,width=0.95\columnwidth]{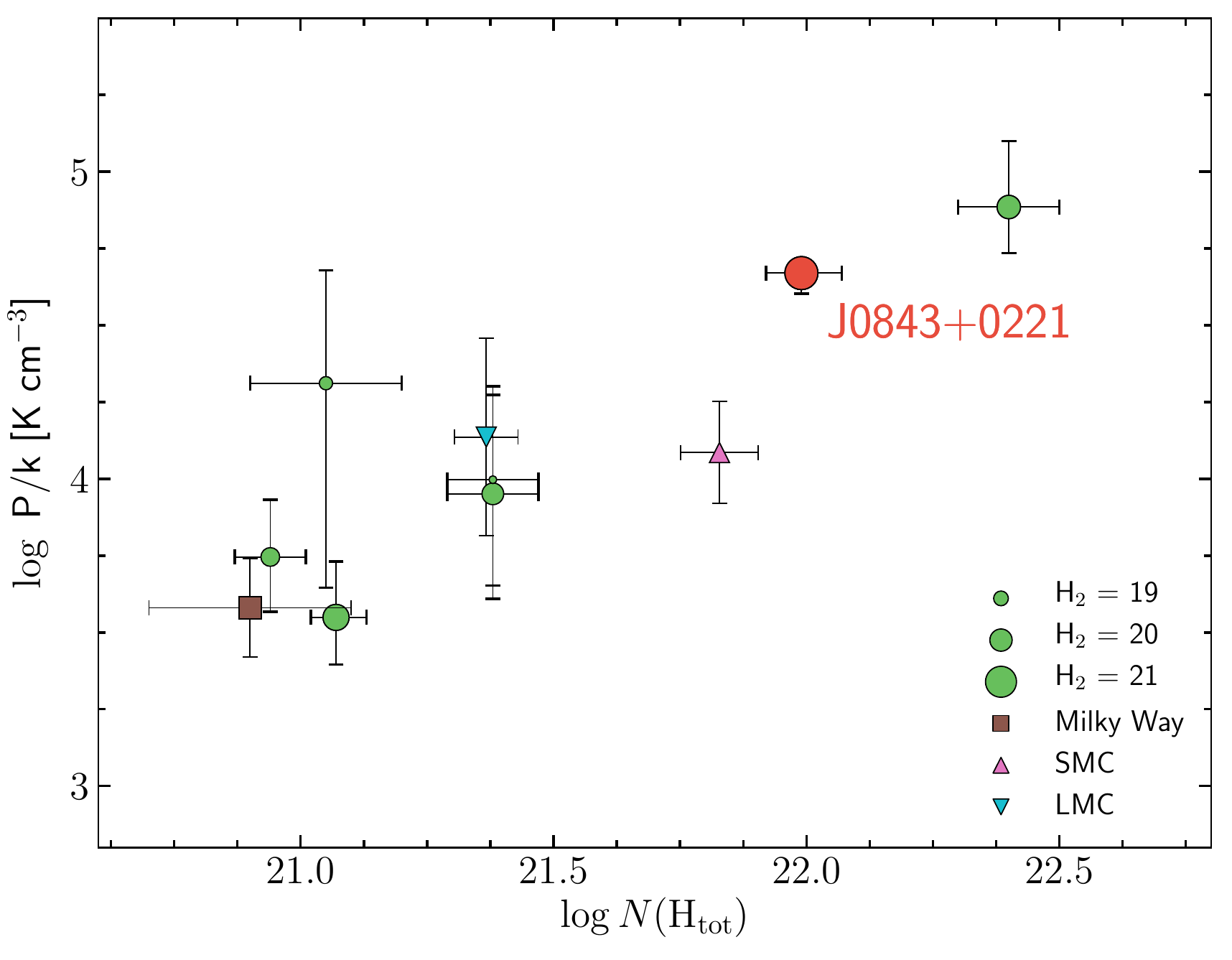}
		\caption{
			Estimated thermal pressure versus total hydrogen column density. Circles represent values derived from the \ion{C}{i} fine-structure levels together with the H$_2$ excitation temperature, T$_{01}$. The size of each circle corresponds to the H$_2$ column density. Squares correspond to the average values found in the Local Group (MW, SMC and LMC). Measurement in the DLA towards \qso\ is consistent with an increase of the thermal pressures with increasing $\log N(\rm H)$.
                }
		\label{H2_Pn}
	\end{center}
\end{figure}

We applied the same calculations as in the previous section to other known high-redshift \ion{C}{i}-bearing systems to derive the pressure estimates shown as green points in Fig.~\ref{H2_Pn}. Note that we assumed the kinetic temperature for each system to equate the $T_{01}$ H$_2$ excitation temperature and that the cloud is fully molecularized. This a rather strong assumption that should in principle be tested for each individual cloud. Since the collisional coefficients for H$_2$ are slightly higher than for \HI, the densities derived should be treated as upper limits and so the derived pressures. We see that most of the known \ion{C}{i}-bearing systems have pressures $P \sim 10^4$ K$\cdot$cm$^{-3}$. Both the system studied here and the other H$_2$ absorption system in the ESDLA towards J\,2140$-$0321 \citep{Noterdaeme2015} have higher pressure. This is consistent with the interpretation that ESDLAs correspond to lines of sight that are statistically closer to the galaxy centre \citep{Noterdaeme2014}, where we expect the pressure to be higher.

\subsection{UV flux}

While the excitation of the high rotational levels of H$_2$ provides an indication of the strength of the UV field, it is not easy to obtain a quantitative estimate due to the complexity of the excitation. A main difficulty is that radiative pumping and self-shielding effects lead to inhomogeneous distribution of the population of different rotational levels of H$_2$ throughout the cloud \citep[see e.g.][]{Abgrall1992}.
High-$J$ levels are dominant at the cloud surface, while low-$J$ levels are mostly found in the shielded interior, where the physical conditions can be different \citep{Balashev2009}. This could explain the increasing Doppler parameter with increasing rotational level, as often observed for H$_2$ \citep{Lacour2005, Noterdaeme2007, Balashev2009}. The effects of geometry, which is likely more complex than the simple slab or spherical that is generally assumed, as well as the effects from turbulence can therefore play important roles.

\begin{figure}
	\begin{center}
		\includegraphics[clip=,width=0.95\hsize]{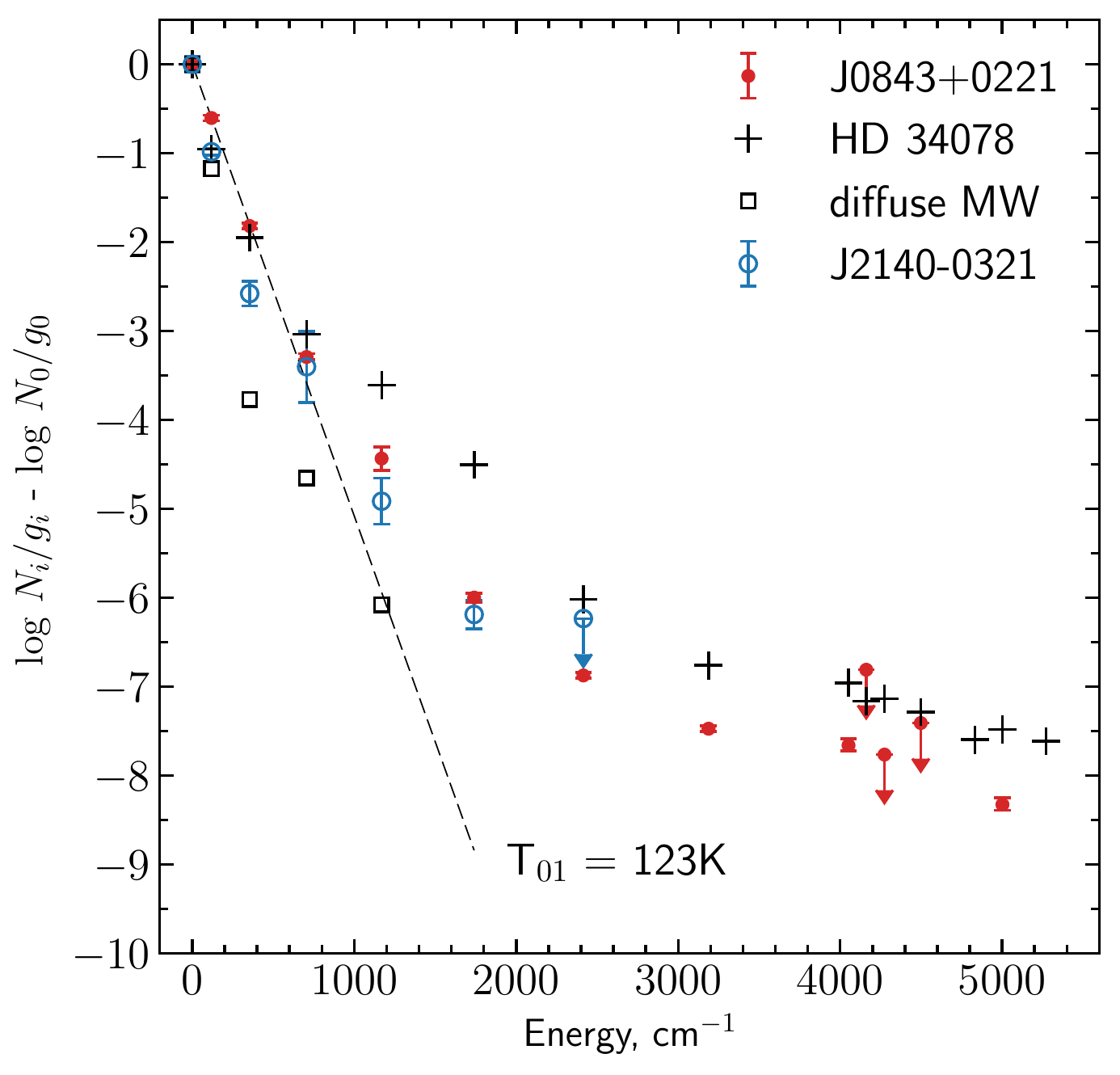}
		\caption{Comparison of H$_2$ rotational excitation in different H$_2$ absorption systems. The black dashed line correspond to the temperature $T_{01}=123$\,K measured using J=0,1 levels of H$_2$ in \qso. { The excitation of H$_2$ suggests the UV flux is enhanced in the H$_2$-bearing medium of the DLA at z=2.7865.}
		}
		\label{H2Exc_total}
	\end{center}
\end{figure}

Keeping this in mind, we consider the H$_2$ column densities in the different rotational levels (see Table \ref{fit_res}) summed over the three components. In Fig.~\ref{H2Exc_total} we compare this 'average' excitation with that of other high-$N($H$_2$) absorption systems: embedded in a typical UV field of the MW \citep{Gry2002}, a highly excited case towards the O9.5V runaway star HD\,34078 \citep{Boisse2005} and the ESDLA at $z=2.34$ towards J\,2140$-$0321 \citep{Noterdaeme2015}. Both high-$z$ DLAs show H$_2$ excitation higher than typical MW values, but lower than towards the OV star. Interestingly, the excitation towards the latter has been successfully modelled using two components, one little affected by the star and contributing to most of the $J=0$ and 1 lines and another one with high density ($n\sim 10^4$~cm$^{-3}$) strongly illuminated by the star. While the situation is unlikely to be such extreme in the DLAs, the excitation diagram indicates an enhanced UV background and larger number densities compared to typical MW values. Indeed, photoionization models also indicate $\log n\sim 2.5-3$ towards J\,2140$-$0321. For \qso\ it can also be seen that the $J=2$ and $J=3$ rotational levels are consistent with the excitation temperature measured from $J=0$ and $J=1$, $T_{01}=123^{+9}_{-8}$\,K, which suggests that the $J=2$ and $J=3$ levels are predominantly populated by collisions in the medium, as expected in strongly self-shielded regions \citep{Abgrall1992}.

An alternative way to constrain the UV field is the \ion{C}{ii*} method \citep{Wolfe2003}. Assuming thermal equilibrium, the heating rate through photoelectric effect on dust grains equates the cooling rate dominated by the [\ion{C}{ii}]158$\mu m$ emission. From the column densities of \ion{C}{ii*}, \HI\ and H$_2$ we find that the cooling rate per hydrogen atom by \ion{C}{ii} fine-structure transition is $\log l_c = -26.0^{+0.2}_{-0.2}$, where $l_c$ is in units of erg\,s$^{-1}$ \citep{Pottasch1979}. This is slightly above the values measured for typical DLAs with $\log N(\HI)<21.5$ \citep{Wolfe2008}. Using the dust-to-gas ratio $\kappa \sim 0.03$ and estimating the grain heating efficiency $\epsilon \sim 0.03$, we find that the UV field is a factor of 10$-$30 higher than the Habbing UV field. This is in qualitative agreement with the high-excitation diagram of H$_2$. This is also in indirect agreement with the low CO/H$_2$ abundance since the photodestruction of CO scales with the UV flux.

On one hand, since our system has a very low dust content, the above method implies that a large UV flux is
required to produce the estimated heating rate. This need not be the case if the heating is mainly due to cosmic rays \citep{Dutta2014,Noterdaeme2017}. In that case, balancing the cooling implies a cosmic ray ionization rate of about $3\times10^{-16}$ s$^{-1}$, i.e. 30 times higher than the canonical Galactic value $10^{-17}$ s$^{-1}$. Note that usually higher values of cosmic ray ionization rate than canonical value are observed in the dense clouds \citep{Dalgarno2006}. On the other hand, it is also expected that the cosmic ray ionization rate correlates with the UV flux, since both arise from star formation activity. Therefore, in the real situation, both dust grain heating and heating by cosmic rays are likely equally important. The factor of 30 derived for the UV flux and cosmic ionization rates alone are therefore upper limits and their actual
values should be roughly a factor of 2 lower. 

\section{Discussion}
\label{discussion}

\subsection{Properties of H$_2$ cloud}
\label{properties}

Using the derived physical conditions, we can discuss some properties of the H$_2$-bearing cloud towards \qso\ and compare them with measurements in the local Universe. In the following, we assume that the H$_2$ cloud is spherical and homogeneous. We again consider the two limiting cases for the molecular fraction, $f=0.3$ and $f=1$. The derived parameters of the cloud are given in Table~\ref{cloud_properties}.

Given the hydrogen column density $N_{\rm H}$ and volumic density, $n$, we find that the cloud size along the line of sight, $l=N_{\rm H}/n$, is between $\sim 1.4$ and $\sim12$ pc (for $f=1$ and $f=0.3$, respectively).
The upper limit is comparable with the characteristic sizes of giant molecular clouds (GMCs) observed in our Galaxy and in the Magellanic Clouds. The mass of the cloud can be estimated as $M = 4/3\pi R^3 \rho$, where $R$ is the radius of the cloud, $\rho = \mu n$, where $\mu$ is the mean mass weight taken to be $2.15 m_{\rm H}$ and $1.23 m_{\rm H}$ ($m_{\rm H}$ is the hydrogen atomic mass) for $f=1$ and $f=0.3$, respectively. Since the impact parameter between the line of sight and the cloud centre is unknown, we use the most probable radius of the cloud $R = 3/4 \cdot l$. We then obtain masses in the range $(1-300)\times 10^2 M_{\odot}$, which corresponds to the lower end of the mass distribution of MW molecular clouds selected from $^{13}$CO emission \citep{RomanDuval2010}. The high allowed values reach the lower limit of $\sim$10$^4$\,M$_{\odot}$, which usually is considered for GMCs \citep{McKee2007}.

We found that the surface density, $\Sigma=M/\pi R^2$ (which in our case is equal to $\mu N_{\rm H}$), is constrained to be in the range $\sim 28-112$~M$_{\odot}$\,pc$^{-2}$ (depending on the adopted molecular fraction).

\begin{figure}
	\begin{center}
		\includegraphics[clip=,width=0.95\columnwidth]{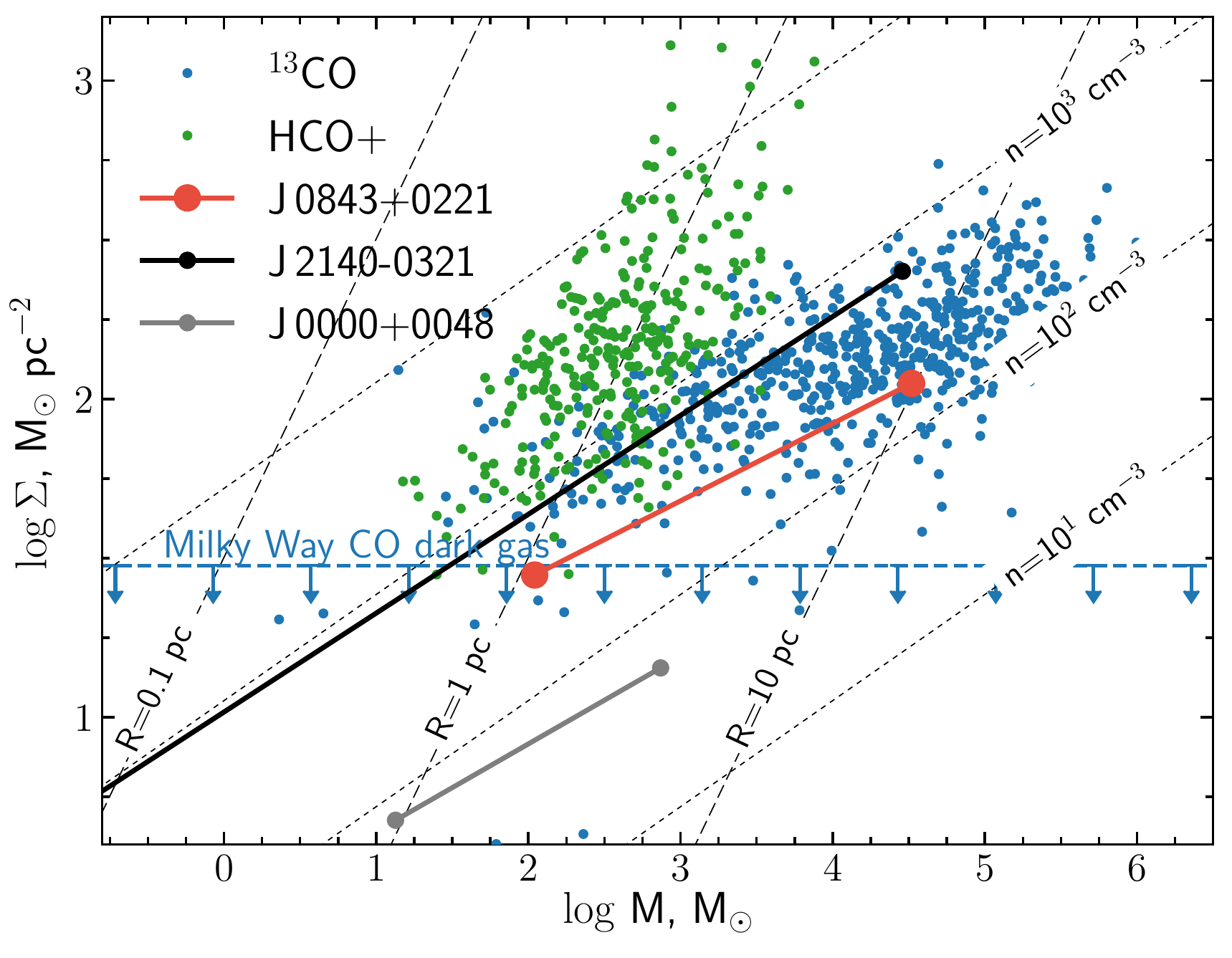}
		\caption{
			Estimates of surface density of cold gas, $\Sigma$, versus mass of the cloud $M$. The blue and green points correspond to measurements of molecular clouds in MW using $^{13}$CO \citep{RomanDuval2010} and HCO$^{+}$ \citep{Barnes2011}. The lines connect the values for extreme cases (only H$_2$ or all H\,{\sc i} associated with H$_2$) for three strong H$_2$-bearing DLAs. The cloud associated with DLA towards \qso\ is most likely similar to $^{13}$CO-selected cloud in the MW.
                }
		\label{M-sigma}
	\end{center}
\end{figure}

In Fig.~\ref{M-sigma} we show the $\Sigma-M$ dependence of molecular clouds. The allowed range for the DLA  towards \qso\ is located slightly below most of the HCO$^{+}$- and $^{13}$CO-selected clouds in the Milky-Way \citep{RomanDuval2010, Barnes2011}. We note in passing that HCO$^{+}$-selected clouds have higher $\Sigma$-values than $^{13}$CO-selected ones for a given mass, due to HCO$^{+}$ tracing high-density clumps in molecular clouds. The observed number density towards \qso\ is comparable to the average value obtained in $^{13}$CO-selected clouds.

While the upper limit of $\Sigma=112$ M$_{\odot}$\,pc$^{-2}$ ($f=0.3$ case) resembles the typical surface densities of MW clouds, the lower limit $\Sigma=28$ M$_{\odot}$\,pc$^{-2}$ ($f=1$ case) corresponds to the lower end of $^{13}$CO-selected clouds in the MW. We note that in the case of the molecular cloud towards $J\,0000+0048$ \citep{Noterdaeme2017}, the surface density remains lower than the MW clouds. In the case of the ESDLA towards $J\,2140-0321$ \citep[$\log N(\HI)=22.4$][]{Noterdaeme2015}, the upper range overlaps with MW points but this corresponds to the extreme case where all \HI\ in the DLA is assumed to be associated with the H$_2$ phase, while modelling suggests that it is not \citep{Noterdaeme2015}. Here, independently of the exact actual molecular fraction, the H$_2$ cloud appears to probe a low-metallicity analogue of $^{13}$CO-selected molecular cloud.

Based on their data, \citet{RomanDuval2010} found that most $^{13}$CO clouds are gravitationally confined with virial parameter $\alpha_{\rm vir}<1$. To estimate the virial parameter, $\alpha_{\rm vir} = 5\sigma^2R/(GM)$, one needs to know the 1D velocity dispersion in the medium, $\sigma$. Using $\sigma\sim3$ km\,s$^{-1}$ from the Doppler parameter of the main \ion{Cl}{i} component, we obtain $\alpha_{\rm vir}$ between $\approx100$ and $3$ for $f=1$ and $f=0.3$, respectively. This means that the cloud is probably not self-gravitating, but pressure confined. This is especially true if we take into account the fact that
$b_{\rm turb}$ measured using low-ionization metals is significantly higher than that based on \ion{Cl}{I}. The range of Doppler parameters suggests that turbulence in the cloud is supersonic with Mach number $>3$. 

\begin{table}
	\centering
	\addtolength{\tabcolsep}{-2pt}
	\setlength\extrarowheight{5pt}
	\caption{Properties of the H$_2$-bearing cloud at z=2.7865.  All values given above are only indicative and do not take into account the uncertainties, which in some cases can be large.      
		\label{cloud_properties}}
	\begin{tabular}{c c c }
		\hline
		\hspace{0.3cm} Parameter \hspace{0.3cm} & \hspace{0.3cm} $f=1$ \hspace{0.3cm} & \hspace{0.3cm} $f=0.3$ \hspace{0.3cm} \\ 
		\hline
		$n$, $10^2$\,cm$^{-3}$ & 3.8 & 2.6 \\
		$P$, $10^4$\,K cm$^{-3}$ & 4.7  & 3.2 \\
		\hline
		$l$, pc     & 1.4 & 12 \\
		$\Sigma$, M$_{\odot}$ pc$^{-2}$  & 28 & 112 \\
		$M$, M$_{\odot}$  & 94 & 2.9$\times10^{4}$ \\
		$\alpha_{\rm vir}$      & 115 & 3 \\
		\hline
		$t_{\rm H_2}$, Myr & 63 & 91 \\
		$t_{\rm ff}$, Myr & 1.8 & 2.7 \\
		$t_{\rm dyn}$, Myr & 0.5 & 4.0 \\
		$t_{\rm cool}$, Myr & 0.1 & 0.1 \\
			\hline
	\end{tabular}
	\addtolength{\tabcolsep}{3pt}
\end{table}

\subsection{Timescales}

The low metallicity and dust-to-gas ratio in the DLA studied here as well as its marked difference with other H$_2$-bearing DLAs call for a discussion of the times-cales of relevant chemical and thermodynamic processes.
Indeed, the rates of these processes usually depend on the metal/dust content, and even for the solar-metallicity gas, the steady-state balance is sometimes questioned. 
 
In the cold ISM, H$_2$ forms only on dust grains. The rate of such H$_2$ formation is relatively slow and therefore the dust content plays a crucial role. This can be estimated as $t = 1/(Rn^2D)$, where $R$ is the formation rate coefficient, $n$ is the number density in the cloud and $D$ is the dust-to-gas ratio compared to the Galaxy one. We assumed that $D$ is proportional to the metallicity measured in \qso\ DLAs, i.e., we
took $D = 1/30$. For the formation rate coefficient, we used the generally adopted value derived from Copernicus observations, $R = 3\times 10^{-17}\, \rm cm^3s^{-1}$. Using our derived number density, we obtain formation time-scale to be $\sim60$ and $\sim 90$ Myr for $f=1$ and $f=0.3$ cases, respectively (see Sect.~\ref{properties}). The dynamical time-scale can be estimated as $t_{\rm dyn} = l / b_{\rm turb}$, where $l$ is the size of the cloud. Using the lower limit on the Doppler parameter $b_{\rm turb} = 3$~km\,s$^{-1}$ from \ion{Cl}{I} we obtained $t_{\rm dyn}= 0.5$ and $4.0$ Myr (for $f=1$ and $f=0.3$, respectively).
The free-fall time can be estimated as $t_{\rm ff} = (3\pi/32/G/\rho)^{1/2}$ and equals $1.8$ and $2.7$ Myr (for $f=1$ and $f=0.3$, respectively). Since the H$_2$ formation timescale is much larger than the dynamical and free fall time, this independently confirms that the H$_2$-bearing cloud is likely pressure confined.
We note that while a high dust content reduces the H$_2$ formation time-scale, the number density can have major importance since $t\propto n^{-2}$. It has two direct consequences: first, in low-metallicity environments, H$_2$ may trace only the densest regions in the patchy medium of DLAs. High number densities will obviously severely reduce the covering fraction of the H$_2$-bearing gas. Secondly, the formation time-scale of H$_2$ can be shortened by supersonic turbulences that can create pockets of compressed gas, where H$_2$ is quickly formed, and consequently redistributed to the adjacent medium.

In Sect.~\ref{T01}, we showed that the $H_2$-bearing gas in the DLA at z=2.7865 has a kinetic temperature $T\sim122$K which is significantly higher than what is expected at such H$_2$ column density. 
We supposed that such high temperature was likely attributed to the low metallicity in the medium. The cooling time can be easily estimated from the cooling rate, $l_c$, derived from \ion{C}{II*}, as $\sim 2 k_{\rm B}T/l_c$, where $k_{\rm B}$ is the Boltzmann constant (the factor of 2 was chosen since the molecular fraction is not well constrained and the correct coefficients are $3/2$ and $5/2$ for monatomic and diatomic gas, respectively). This indicates a cooling time of about 0.1~Myr, which is at least an order shorter than the time-scales mentioned above. Hence the cloud is likely in thermal equilibrium and the high kinetic temperature is indeed due to the low metallicity and not to the formation of H$_2$ in a shock-compressed medium.

\subsection{CO content, $X_{\rm CO}$ and CO-dark gas}

In spite of extremely high H$_2$ column density, we did not detect CO. This is not surprising given the facts that we measured a low associated extinction and possibly an enhanced UV flux: CO is easily photodestructed by UV photons in the diffuse medium and therefore at least moderate level of extinction $A_V\gtrsim 1$ is required for this molecule to survive. Models indicate that the abundance of CO scales with the metallicity, but there are only scarce measurements at low metallicities \citep[e.g.][]{Shi2015}. Our upper limit on $N($CO) is therefore very valuable since it provides a stringent limit on $X_{\rm CO}$ at $Z \sim 0.03 Z_{\odot}$ for the first time.
	
\begin{figure}
	\centering
	\includegraphics[clip=,width=0.95\hsize]{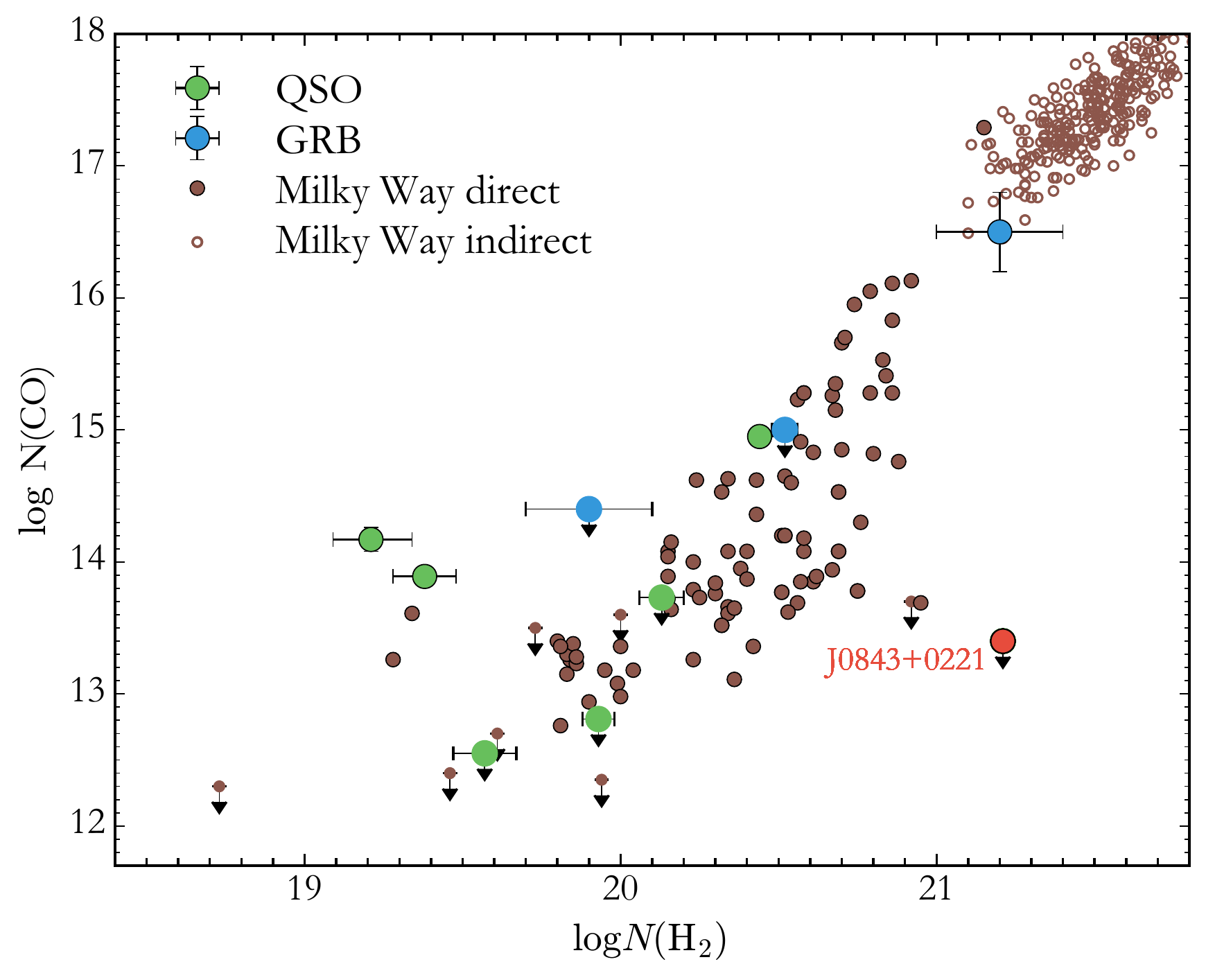} 
	\caption{
		Dependence of $N({\rm CO})$ to $N({\rm H}_2)$ column densities for high-redshift DLAs detected in quasar and GRB afterglow (green and blue points, respectively). The open and filled brown points correspond to direct \citep{Sheffer2008, Burgh2010} and indirect \citep{Federman1990} H$_2$ measurements in our Galaxy, respectively. The upper limit on $N({\rm CO})$ in the H$_2$ absorption system at z=2.7865 studied here is shown by a red circle and is a factor of 10$^3$ lower than values measured for similar H$_2$ column densities in the MW.}
	\label{COH2}
\end{figure}

Fig.~\ref{COH2} shows the dependence of the CO column density on that of H$_2$ measured in both the MW and high-redshift ISM. Values for high-$z$ DLAs are from \citet{Srianand2008,Noterdaeme2010,Noterdaeme2017} for quasar absorption systems and \citet{Prochaska2009b} for the GRB-DLA\footnote{A few additional CO measurements  are available \citep{Noterdaeme2009a,Noterdaeme2011,Ma2015} for quasar DLAs but their lower redshifts prohibit direct H$_2$ measurements from ground-based telescopes.}.
We also derived upper limits on the total CO column density in known strong H$_2$ absorption systems, using the covered A-X bands in the archival spectra and assuming that the population of CO rotational levels is dominated by the CMB excitation. We found $N({\rm CO})<12.55$ in the DLA towards Q\,$1232+082$ with $\log N({\rm H}_2)=19.57\pm0.10$ and $N({\rm CO})<12.81$ in the DLA towards Q\,$0812+3208$ with $\log N({\rm H}_2)=19.93\pm0.05$. A correlation between $N($CO) and $N($H$_2$) is naturally expected as we probe larger column densities of molecular gas  but there is interesting deviation around the main trend. CO detections in DLAs tend to probe high CO/H$_2$ values. This could be due to selection effects since the measurements are close to the typical CO detection limits. In the case of detections, the high CO/H$_2$ ratio could result from a combination of small dust grains and high cosmic ray ionization rate  \citep[see][]{Shaw2008,Shaw2016,Noterdaeme2017}. The case of the extreme H$_2$ absorber studied here is striking: our limit on $N($CO) is 10$^3$ times lower than what the MW trend suggests or than the value measured in the GRB-DLA by \citet{Prochaska2009b}. This can be attributed to the much lower gas-phase abundances of C and O in the DLA towards \qso\ together with the low dust content, which is crucial for CO since dust provides shielding from the UV field.

\begin{figure}
	\centering
	\includegraphics[clip=,width=0.95\hsize]{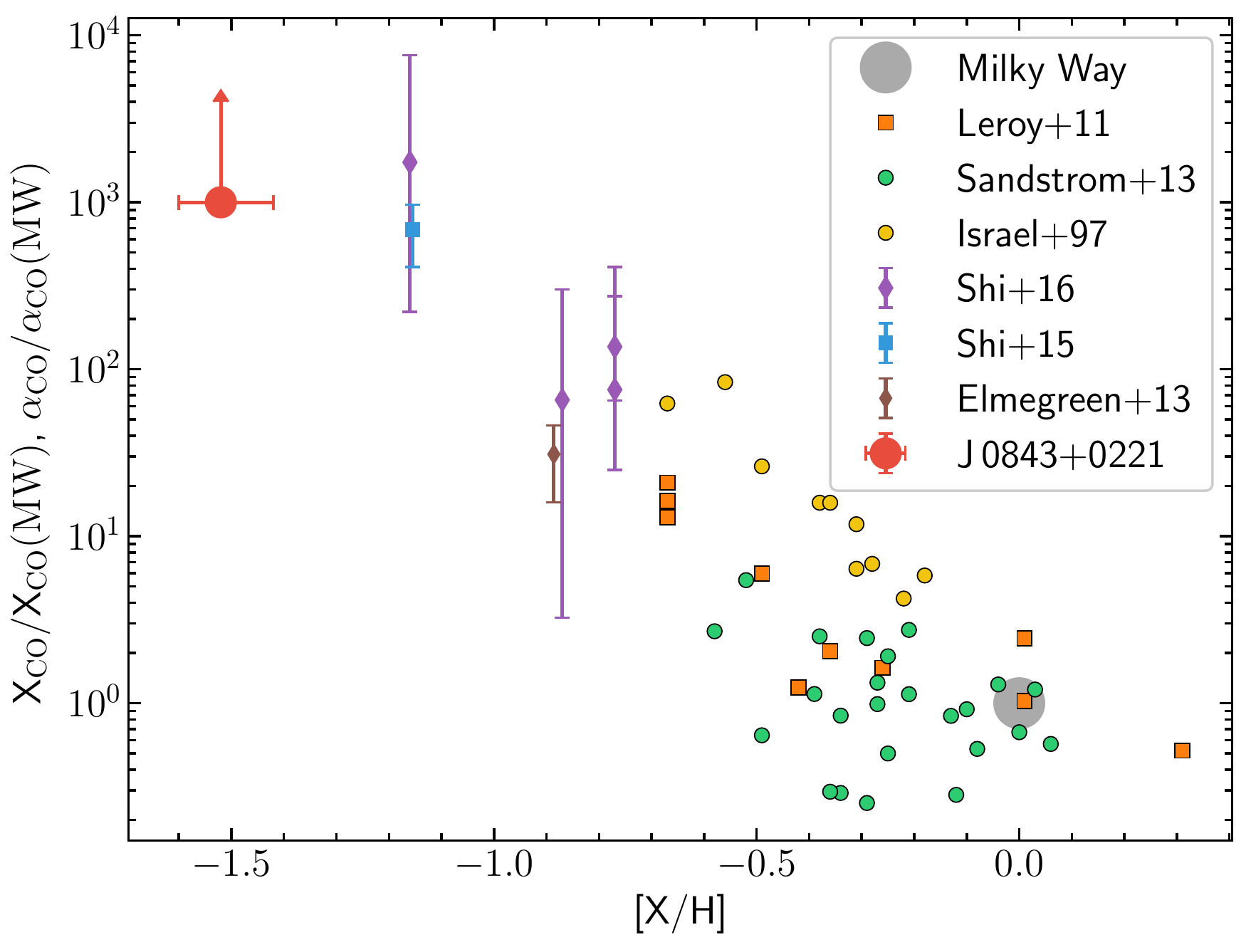} 
	\caption{
		 Measurements of $X_{\rm CO}$ and $\alpha_{\rm CO}$ conversion factors relative to adopted MW values \citep{Bolatto2013} as a function of metallicity. The data for the local measurements of $\alpha_{\rm CO}$ are from \citealt{Israel1997, Leroy2011, Sandstrom2013, Elmegreen2013, Shi2015, Shi2016}. }
	\label{XCO}
\end{figure}

We calculated the lower limit on the $X_{\rm CO}$ conversion factor for the optically thin regime \citep[details can be found in][]{Bolatto2013} to be $X_{\rm CO}\gtrsim 2-6 \times 10^{23}\, \rm cm^{-2}/(km/s\,K)$, depending on the CO excitation temperature that must be in the range between the CMB temperature at $z=2.7865$ and the kinetic temperature (from $T_{01}$). This value of $X_{\rm CO}$ is $\gtrsim1000-3000$ times higher than the adopted MW value. Given this very high $X_{\rm CO}$ value (or low  CO/H$_2$ ratio), we can state that our high-$z$ molecular cloud corresponds to CO-dark gas.\footnote{We note that, in the MW, the 'CO-dark gas' (or 'CO-faint gas', see \citealt{Wolfire2010,Bolatto2013}) is usually estimated to have $\Sigma < 30$ M$_{\odot}$\,pc$^{-2}$. This threshold value in the MW corresponds to $\log N(\rm CO) \sim 16.5$ (see Fig.~\ref{COH2})} 
Such a high $X_{\rm CO}$ conversion factor is expected at low metallicity and enhanced UV flux in the medium, and any observational constraint is valuable. Fig.~\ref{XCO} summarizes the measurements of CO/H$_2$ conversion factors ($X_{\rm CO}$, $\alpha_{\rm CO}$) as a function of metallicity in different environments.  While we are only able to put a lower limit on $X_{CO}$ that is similar to previous measurements at low metallicities, it is the first time a useful constraint is put at such a low metallicity (twice lower than any previous measurement in the MW).    

\subsection{Star formation rates}

Following the formalism of \citet{Wolfe2008} and the estimates on the UV flux, we estimate the surface
star formation rate, $\Sigma_{\rm SFR}$, in the DLA towards \qso\ to be about  $2\times10^{-2}$\,M$_{\odot}$\,yr$^{-1}$\,kpc$^{-2}$, if stars are forming {\it in situ}. This value is comparable with the values of $\Sigma_{\rm SFR}$ for the measured $\Sigma_{\rm H2}$ in local galaxies with kpc-resolution emission observations \citep{Bigiel2011}. This supports the idea that the H$_2$-bearing system in \qso\ probes a star-forming medium, similar to the GMCs in the local Universe.

Detecting more direct markers of star formation such as the nebular emission lines would be very valuable \citep[see e.g.][]{Krogager2017}. Here, only the Ly-$\alpha$ line is covered by our UVES spectrum. This line can in principle be detected in the bottom of the damped Ly-$\alpha$ absorption that removes the flux of the quasar over a significant wavelength range \citep[as in][]{Noterdaeme2012a}. However, we did not find any significant Ly-$\alpha$ emission in our spectrum. To estimate a limit on the Ly-$\alpha$ flux, we rebinned and coscaled the UVES spectrum with the flux-calibrated SDSS spectrum and obtained $F_{\rm Ly\alpha} < 0.8 \times 10^{-17} \rm erg\, cm^{-2}\, s^{-1}$ (and Ly-$\alpha$ luminosity $< 5.7\times10^{41} \rm erg\, s^{-1}$. Assuming no dust correction, this upper limit translates \citep[see e.g.][]{Rahmani2010} to $\rm SFR < 4 $ M$_{\odot}\rm \, yr^{-1}$ for {\sl in situ} star formation activity.  

\section{Conclusion}
\label{conclusion}
\noindent

We presented the analysis of the very high H$_2$ column density absorption system at $z_{\rm abs}=2.7865$ identified in the spectrum of quasar \qso. We measured the total H$_2$ column density to be $\log N ({\rm H}_2) = 21.21\pm0.02$, which is a factor of 6 higher than what has been previously detected in intervening DLAs towards quasars. The relative population of $J=0,1$ levels of H$_2$ indicates a kinetic temperature in the H$_2$-bearing medium of $T=123^{+9}_{-8}$\,K, which is at least twice higher than what is expected for such large H$_2$ column densities from extrapolating the trend seen in other measurements. This is probably due to the low metallicity of the medium. We find that metallicity in this ESDLA ($\log N(\HI) \sim 21.82^{+0.11}_{-0.11}$) is one of the lowest, $\rm [Zn/H] \sim -1.5$, among H$_2$-bearing DLAs. Although the measured metallicity is typical for DLAs at similar redshift, the depletion level in the DLA, $\rm [Fe/Zn]\sim -1$, is found to be significantly higher than in non H$_2$-bearing DLAs at the same metallicity.
This supports the idea that some amount of dust is required for the presence of H$_2$. Even though the depletion and the total hydrogen column density are high, the dust content is still not enough to produce any significant associated extinction. We measure $A_V=0.09\pm0.10$, i.e within the systematic uncertainties of the measurements. This is 30 times lower than what is expected using the average MW $A_V/N(\rm H)$ ratio, but scales exactly with the measured metallicity. 

The low metallicity implies a very low abundance of CO. We measure an upper limit on the column density of CO, $\log N (\rm CO) < 13.35$ that  translates to a lower limit on $X_{\rm CO}$ factor $>2 \times 10^{23}$ cm$^{-2}$ / (km\,s$^{-1}$\,K). This is $>$10$^3$ times higher than the adopted MW value. While an increase of $X_{\rm CO}$ with decreasing metallicity is expected, our system has the lowest metallicity in which $X_{\rm CO}$ is constrained. The low CO and high H$_2$ content indicate that this system represents 'CO-dark/faint' gas, whose abundance in different galaxies is currently a matter of debate now \citep{Grenier2005, Pineda2013, Smith2014}.

The detection of fine-structure levels of \ion{C}{i}, \ion{C}{ii}, \ion{Si}{ii} and the excitation of HD molecules allows us to constrain the number density in the cold gas to be $\sim260-380$ cm$^{-3}$, depending on the adopted molecular fraction. Taking into account our estimate of the kinetic temperature, we found a relatively high thermal pressure in this system, $\sim 3-5 \times 10^4$ K\,cm$^{-3}$, which is higher than typical pressures measured in the diffuse ISM at $z=0$. Our measurement confirms the trend of increasing
thermal pressure with increasing hydrogen column density. This is in line with the idea that high column density systems preferably probe gas at small impact parameters from the associated galaxy \citep{Noterdaeme2014}. The observed column density of \ion{C}{II*} implies a cooling rate by
[\ion{C}{ii}]\,158\,$\mu$m to be $l_c \sim 10^{-26}$~erg\,s$^{-1}$, which is slightly higher than what is usually observed in DLAs. This, together with the excitation diagram of H$_2$ rotational levels and the low CO abundance, indicates a strong UV field, roughly a few tens times higher than the Habing field.

The molecular fraction in the H$_2$-bearing medium can be in range between 0.3 and 1. Assuming homogeneous density, we estimate the size of this cloud to be between $\sim$1 and 10 pc along the line of sight. The cloud is most probably pressure confined, based on the estimation of the virial parameter and characteristic time-scales. Assuming spherical shape of the cloud, we derive the total gas mass between $(1-300)\times10^4$\,M$_{\odot}$, while the higher range reaches the commonly used threshold for GMC. The corresponding surface density is between $\sim$30 and 110 M$_{\odot}$\,pc$^{-2}$. This is very similar to the $^{13}$CO-selected clouds studied in our Galaxy. This means that for the first time in quasar absorption line studies, we confidently probe a starforming molecular cloud similar to GMCs in the MW.

\vspace{2mm}{\footnotesize {\rm Acknowledgments:}
	
  Based on observations collected at the European Organisation for Astronomical Research in the Southern Hemishpere under ESO programme 092.A.0345(A). We are very grateful to the anonymous referee for the detailed and careful reading our manuscript and many useful comments. This work was supported by the Russian Science Foundation grant 14-12-00955. SB thanks the French Government, Campus France and the Institut Lagrange de Paris for support during his visits at IAP. SB and RS thank IAP for hospitality and support during the time part of this work was done. PN, PPJ and RS gratefully acknowledge support from Indo-French Centre for the Promotion of Advanced Research under project No. 5504-B.   PN acknowledges support from the 'Programme National de Cosmologie and Galaxies' (PNCG) funded by CNRS/INSU-IN2P3-INP, CEA and CNES, France.

\def\aj{AJ}%
\def\actaa{Acta Astron.}%
\def\araa{ARA\&A}%
\def\apj{ApJ}%
\def\apjl{ApJ}%
\def\apjs{ApJS}%
\def\ao{Appl.~Opt.}%
\def\apss{Ap\&SS}%
\def\aap{A\&A}%
\def\aapr{A\&A~Rev.}%
\def\aaps{A\&AS}%
\def\azh{AZh}%
\def\baas{BAAS}%
\def\bac{Bull. astr. Inst. Czechosl.}%
\def\caa{Chinese Astron. Astrophys.}%
\def\cjaa{Chinese J. Astron. Astrophys.}%
\def\icarus{Icarus}%
\def\jcap{J. Cosmology Astropart. Phys.}%
\def\jrasc{JRASC}%
\def\mnras{MNRAS}%
\def\memras{MmRAS}%
\def\na{New A}%
\def\nar{New A Rev.}%
\def\pasa{PASA}%
\def\pra{Phys.~Rev.~A}%
\def\prb{Phys.~Rev.~B}%
\def\prc{Phys.~Rev.~C}%
\def\prd{Phys.~Rev.~D}%
\def\pre{Phys.~Rev.~E}%
\def\prl{Phys.~Rev.~Lett.}%
\def\pasp{PASP}%
\def\pasj{PASJ}%
\def\qjras{QJRAS}%
\def\rmxaa{Rev. Mexicana Astron. Astrofis.}%
\def\skytel{S\&T}%
\def\solphys{Sol.~Phys.}%
\def\sovast{Soviet~Ast.}%
\def\ssr{Space~Sci.~Rev.}%
\def\zap{ZAp}%
\def\nat{Nature}%
\def\iaucirc{IAU~Circ.}%
\def\aplett{Astrophys.~Lett.}%
\def\apspr{Astrophys.~Space~Phys.~Res.}%
\def\bain{Bull.~Astron.~Inst.~Netherlands}%
\def\fcp{Fund.~Cosmic~Phys.}%
\def\gca{Geochim.~Cosmochim.~Acta}%
\def\grl{Geophys.~Res.~Lett.}%
\def\jcp{J.~Chem.~Phys.}%
\def\jgr{J.~Geophys.~Res.}%
\def\jqsrt{J.~Quant.~Spec.~Radiat.~Transf.}%
\def\memsai{Mem.~Soc.~Astron.~Italiana}%
\def\nphysa{Nucl.~Phys.~A}%
\def\physrep{Phys.~Rep.}%
\def\physscr{Phys.~Scr}%
\def\planss{Planet.~Space~Sci.}%
\def\procspie{Proc.~SPIE}%
\let\astap=\aap
\let\apjlett=\apjl
\let\apjsupp=\apjs
\let\applopt=\ao

\bibliographystyle{mn2e}
\bibliography{H2}

\appendix

\section{Ionization structure, fitting parameters and details on H$_2$ J$\ge$2 rotational levels}
\label{metal_appendix}

Fig.~\ref{ioniz} presents the ionization structure of the DLA at $z=2.7865$, which is found to be quite typical of what is seen in other DLAs. Table~\ref{fit_res} presents the best-fitting parameters to the absorption lines in the neutral (atomic and molecular) phases. 

\begin{figure}
	\begin{center}
		\includegraphics[clip=,width=1.0\hsize]{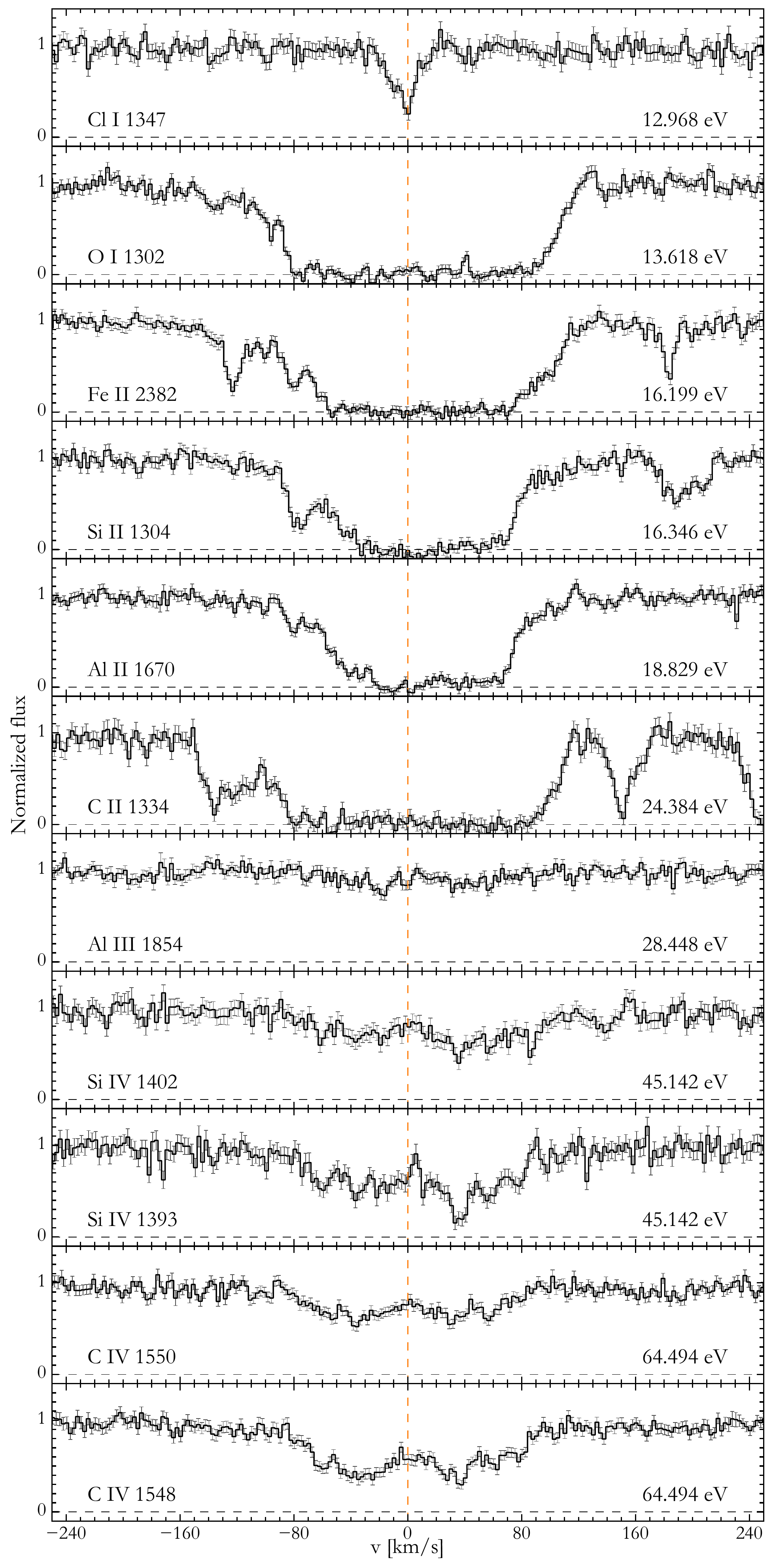}
		\caption{Comparison of the line profiles of different ionization species to probe the ionization structure of the DLA at z=2.7865. We keep only lines of interest, not used in the fit. The orange vertical line corresponds to redshift z=2.786574 (component B). The ionization energy of the species is given in each panel.}
		\label{ioniz}
	\end{center}
\end{figure}

\begin{table*}
	\centering
	\addtolength{\tabcolsep}{-3pt}
	\setlength\extrarowheight{3pt}
	\caption{Fitting results. Each section of Table delimited by horizontal lines contains species that were fitted individually.      
		\label{fit_res}}
	\begin{tabular}{lccccccccccc}
		\hline \hline
		& \multicolumn{3}{c}{Component $A$} & \multicolumn{3}{c}{Component $B$} & \multicolumn{3}{c}{Component $C$} & total \\
		species & $z$ & $\log N$ & $b,$ km s$^{-1}$ & $z$ & $\log N$ & $b,$ km s$^{-1}$ & $z$ & $\log N$ & $b,$ km s$^{-1}$ & $\log N$ \\
		\hline
		\ion{Cl}{I} & 2.786459(33) & $13.21^{+0.18}_{-0.11}$ & $7.9^{+4.3}_{-1.3}$ & 2.786582(9) & $13.35^{+0.28}_{-0.16}$ & $2.5^{+1.7}_{-1.2}$ & 2.786720(55) & $12.62^{+0.26}_{-0.30}$ & $3.9^{+8.5}_{-2.1}$ & $13.63^{+0.20}_{-0.05}$ \\
		\hline 
		\HI & \multicolumn{9}{c}{One component fit with $z=2.786540(150)$} & $21.82\pm0.11$ \\
		\hline 
		H$_2$, J=0  & \multicolumn{9}{c}{One component fit with $z=2.786570(14)$}  & $20.71\pm0.02$ \\
		H$_2$, J=1  & \multicolumn{9}{c}{''}& $21.06\pm0.02$ \\
		H$_2$, total & \multicolumn{9}{c}{}& $21.21\pm0.02$ \\
		H, total & \multicolumn{9}{c}{}& $21.99^{+0.08}_{-0.07}$ \\
		\hline
		H$_2$, J=2  & 2.786459$^a$ & $19.47^{+0.03}_{-0.05}$ & $8.8^{+0.1}_{-0.3}$ &  2.786582$^a$ & $19.10^{+0.04}_{-0.19}$ & $6.1^{+0.2}_{-0.2}$ & 2.786720$^a$ & $16.00^{+0.30}_{-1.70}$ & $9.4^{+0.9}_{-0.4}$ & $19.59^{+0.02}_{-0.02}$ \\
		H$_2$, J=3  & '' & $17.67^{+0.09}_{-0.20}$ & '' & '' & $18.71^{+0.02}_{-0.03}$ & '' & '' &  $13.29^{+0.77}_{-0.14}$ & '' & $18.74^{+0.02}_{-0.03}$ \\
		H$_2$, J=4  & '' & $16.44^{+0.08}_{-0.05}$ & '' & '' & $17.14^{+0.15}_{-0.13}$ & '' & '' &  $14.80^{+0.08}_{-0.05}$ & '' & $17.23^{+0.13}_{-0.13}$ \\
		H$_2$, J=5  & '' & $15.89^{+0.05}_{-0.05}$ & '' & '' & $15.85^{+0.10}_{-0.05}$ & '' & '' &  $15.30^{+0.06}_{-0.04}$ & '' & $16.23^{+0.05}_{-0.04}$ \\
		H$_2$, J=6  & '' & $14.76^{+0.02}_{-0.03}$ & '' & '' & $13.80^{+0.15}_{-0.38}$ & '' & '' &  $14.45^{+0.04}_{-0.05}$ & '' & $14.95^{+0.02}_{-0.02}$ \\
		H$_2$, J=7  & '' & $14.57^{+0.05}_{-0.02}$ & '' & '' & $13.85^{+0.18}_{-0.41}$  & '' & '' &  $14.54^{+0.04}_{-0.05}$ & '' & $14.89^{+0.02}_{-0.02}$ \\
		H$_2$, J=8  & '' & $14.04^{+0.07}_{-0.07}$ & '' & '' & $13.28^{+0.41}_{-0.49}$ & '' & '' &  $13.88^{+0.09}_{-0.18}$ & '' & $14.28^{+0.06}_{-0.07}$ \\
		H$_2$, J=9  & '' & $13.85^{+0.12}_{-0.09}$ & '' & '' & $13.70^{+0.08}_{-0.25}$ & '' & '' &  $13.38^{+0.22}_{-0.20}$ & '' & $14.14^{+0.06}_{-0.07}$ \\
		H$_2$, J=10  & '' & $<13.3$ & '' & '' & $<13.4$ & '' & '' &  $<12.9$ & '' & $<13.6$ \\
		H$_2$, J=0, $\nu=1$ & '' & $<13.7$ & '' & '' & $<13.4$ & '' & '' & $<13.4$ & '' & $<13.9$ \\
		H$_2$, J=1, $\nu=1$ & '' & $<13.7$ & ''  & '' & $<13.6$ & '' & '' & $<13.5$ & '' & $<13.9$ \\		
		H$_2$, J=2, $\nu=1$ & '' & $<13.7$ & '' & '' & $<13.5$ & '' & '' & $<13.7$ & '' & $<14.0$ \\
		\hline 
		HD, J=0  & 2.786459$^a$ & $15.15^{+0.27}_{-0.12}$ & $6.4^{+0.5}_{-0.6}$ & 2.786582$^a$ & $17.34^{+0.13}_{-0.38}$ &  $1.5^{+0.6}_{-0.4}$ & 2.786582$^a$ & $<14.0$ & & $17.34^{+0.13}_{-0.37}$ \\
		HD, J=1  & '' & $14.99^{+0.03}_{-0.05}$ & '' & '' & $15.81^{+0.75}_{-0.66}$ & '' & '' & $<14.1$ & & $15.87^{+0.72}_{-0.49}$ \\
		HD, total & \multicolumn{9}{c}{}& $17.35^{+0.15}_{-0.34}$ \\
		\hline
		CO & 2.786459$^a$ & $<13.1$ & & 2.786582$^a$ & $<13.0$ & & & & & $<13.4$ \\
		\hline
		\ion{C}{I} & 2.786435(9) & $13.09^{+0.05}_{-0.04}$ & $4.7^{+1.2}_{-0.7}$ & 2.786578(4) & $13.33^{+0.05}_{-0.03}$ & $3.6^{+0.4}_{-0.6}$ & & & & $13.53^{+0.04}_{-0.02}$ \\
		\ion{C}{I}$^*$ & '' & $13.13^{+0.08}_{-0.05}$ & '' & '' & $13.44^{+0.02}_{-0.05}$ & '' & & & & $13.61^{+0.02}_{-0.02}$ \\
		\ion{C}{I}$^{**}$ & ' & $12.90^{+0.10}_{-0.10}$ & '' & '' & $13.14^{+0.04}_{-0.04}$ & '' & & & & $13.34^{+0.04}_{-0.03}$ \\
		\hline
		$b_{\rm turb}^{c}$, km s$^{-1}$ & & & $4.2^{+1.1}_{-1.1}$ & & & $9.1^{+0.6}_{-0.5}$ & & & $6.8^{+2.8}_{-1.8}$ & & \\
		$T^{c}$, 10$^4$\,K & & & $0.8^{+1.2}_{-0.6}$ & & & $1.1^{+0.7}_{-0.7}$ & & & $1.2^{+0.7}_{-0.7}$ &  & \\
		\ion{Mg}{II} & 2.786377(6) & $14.67^{+0.48}_{-1.55}$ & & 2.786555(4) & $15.91^{+0.03}_{-0.05}$ & & 2.786775(3) & $12.72^{+1.91}_{-0.48}$ & & $15.94^{+0.06}_{-0.05}$ \\
		\ion{Si}{II} & '' & $14.94^{+0.07}_{-0.13}$ & & '' & $15.62^{+0.04}_{-0.06}$ & & '' & $14.90^{+0.14}_{-0.14}$ & & $15.77^{+0.04}_{-0.05}$  \\
		\ion{S}{II} & '' & $13.67^{+0.34}_{-0.13}$ & & '' & $15.51^{+0.04}_{-0.02}$ & & '' & $14.59^{+0.23}_{-0.12}$ & & $15.57^{+0.05}_{-0.02}$   \\
		\ion{Ti}{II} & '' & $12.03^{+0.18}_{-0.00}$ & & '' & $12.84^{+0.07}_{-0.07}$ & & '' & $12.24^{+0.24}_{-0.78}$ & & $12.98^{+0.07}_{-0.07}$  \\
		\ion{Cr}{II} & '' & $12.52^{+0.12}_{-0.16}$ & & '' & $13.08^{+0.05}_{-0.05}$ & & '' & $12.24^{+0.16}_{-0.94}$ & & $13.23^{+0.05}_{-0.06}$  \\
		\ion{Mn}{II} & '' & $12.32^{+0.26}_{-0.21}$ & & '' & $13.60^{+0.07}_{-0.06}$ & & '' & $12.64^{+0.21}_{-0.48}$ & & $13.67^{+0.07}_{-0.06}$  \\
		\ion{Fe}{II} & '' & $14.27^{+0.07}_{-0.12}$ & & '' & $14.80^{+0.05}_{-0.04}$ & & '' & $14.04^{+0.16}_{-0.32}$ & & $14.96^{+0.04}_{-0.04}$  \\
		\ion{Ni}{II} & '' & $12.91^{+0.11}_{-0.08}$ & & '' & $13.50^{+0.04}_{-0.06}$ & & '' & $12.97^{+0.15}_{-0.15}$ & & $13.69^{+0.05}_{-0.05}$  \\
		\ion{Zn}{II} & '' & $12.12^{+0.10}_{-0.22}$ & & '' & $12.91^{+0.04}_{-0.05}$ & & '' & $12.12^{+0.06}_{-0.31}$ & & $13.03^{+0.03}_{-0.06}$  \\
		[0.2cm]
		\hline
		\ion{P}{ii} & & & & & & & & & & $<13.72$ \\
		\ion{Si}{ii*} & 2.786377$^{d}$ & $11.32^{+0.14}_{-0.21}$ & & 2.786555$^{d}$ & $11.62^{+0.08}_{-0.14}$ & & & & &  $11.80^{+0.07}_{-0.11}$ \\
		\ion{C}{ii*} & 2.786377$^{d}$ & $<13.6$ & & 2.786555$^{d}$ & $15.55^{+0.16}_{-0.18}$ & & 2.786775$^{d}$ & $<13.4$ & & \\
                \hline
	\end{tabular}
	\addtolength{\tabcolsep}{3pt}
	\begin{tablenotes}
		\item {$^{a}$ The redshifts of the components A,B and C were fixed during the fit and correspond to the best-fitting values from fit to \ClI\ lines (see the text).}
		\item {$^{b}$ To fit \HI\ and H$_2$ J=0, 1 lines we used one-component model, due to high saturation of the absorption lines.}
		\item {$^{c}$ To fit low-ionization metal absorption lines we tied $b$ parameters using microturbulence Doppler parameter, $b_{\rm turb}$, and kinetic temperature, $T$, using standard relation (see text). }
		\item {$^{d}$ The redshifts of the components A,B and C were fixed during the fit and correspond to the best-fitting values from fit to low-ionization metal lines (see the text).}
	\end{tablenotes}
\end{table*}

\label{H2_appendix}

We used the three components derived from \ClI\ with fixed
redshifts (given in Table~\ref{fit_res}) to fit the $J\ge 2$ lines of H$_2$ and obtain column densities in each component after
locally adjusting the continuum level using spline interpolation. For each component we tied the Doppler parameters
($b$) of the different rotational levels. The results of the analysis are presented in Table~\ref{fit_res} and the profiles of fitted H$_2$
lines are shown in Fig.~\ref{H2high}.  Since it has been shown by different authors that the Doppler parameters of H$_2$ lines tend
to increase with the rotational level \citep{Lacour2005,Noterdaeme2007,Balashev2009}, we also performed a fit with independent $b$-values for each rotational
level. We indeed found the increase of Doppler parameter with $J$ for the two stronger components of H$_2$ absorption system.
Nevertheless, the derived column densities remain consistent with the fit using tied Doppler parameters, although the uncertainties become
much larger due to this additional freedom.  

\begin{figure*}
	\begin{center}
		\includegraphics[clip=,width=1.0\textwidth]{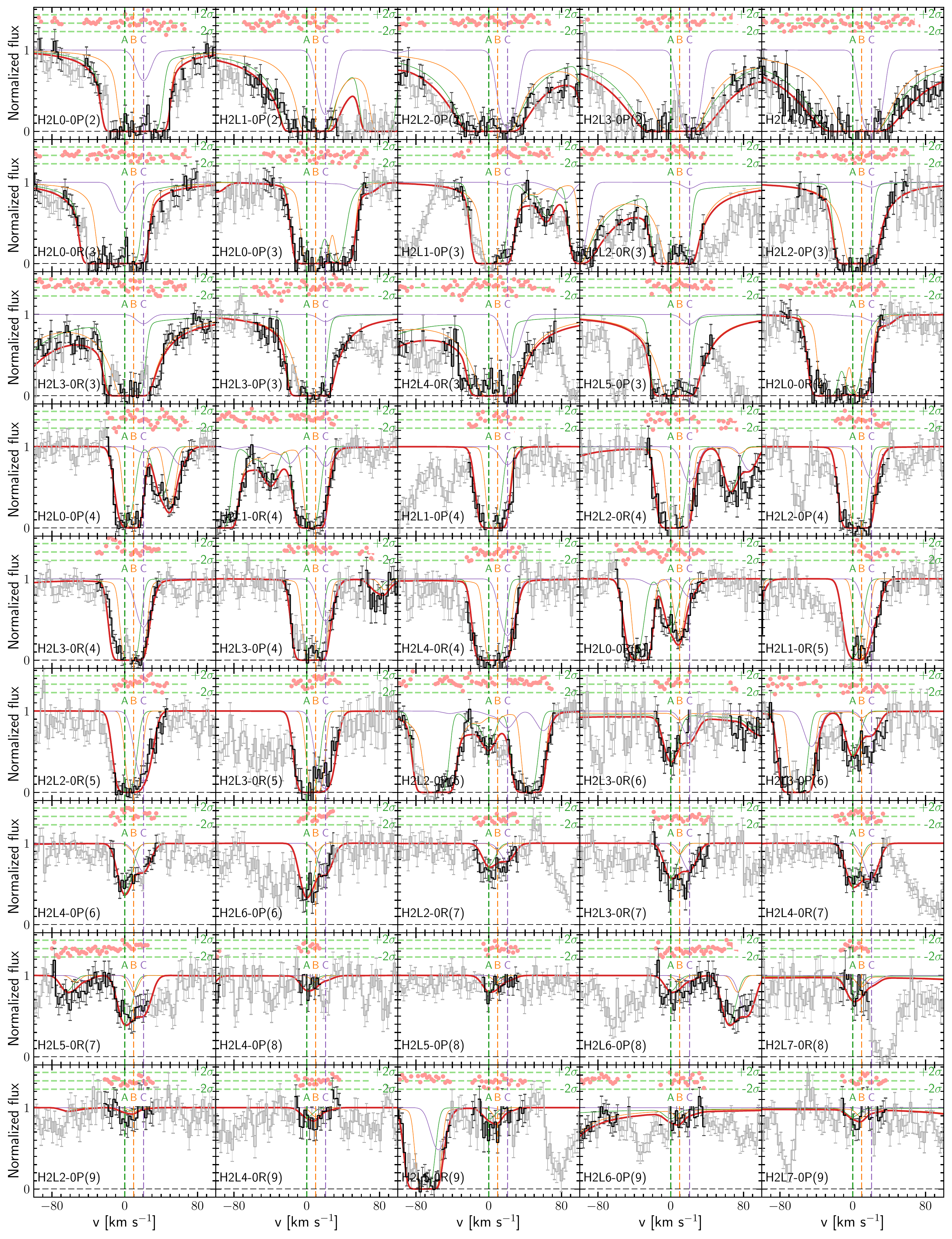}
		\caption{Fit to the H$_2$ absorption lines. Lines of transitions from rotational levels $J=2$ to 9 of the $\nu=0$ vibrational level. Lines are as in Fig.~\ref{Chlorine}.}
		\label{H2high}
	\end{center}
\end{figure*}

The excitation diagrams of H$_2$ for the components $A$, $B$ and $C$ are shown in Fig.~\ref{H2Exc}. The $J=0,1$ points represent the total column densities in the system and so are strict upper limits for the individual components. Focusing on $J\ge 2$ levels,
it appears that the components A and B have similar diagrams, with slightly larger column densities for the component B but slightly higher excitation
temperature for the component A. The component C is the weakest of the three, but apparently has much higher excitation, as suggested by the shallow
slope between $J \ge 2$ levels. Interestingly, this component can therefore contribute significantly to the total high-$J$ column densities,
but little to the overall H$_2$ column density, which mainly arises from the components A and B.

\begin{figure}
	\begin{center}
		\includegraphics[clip=,width=0.95\hsize]{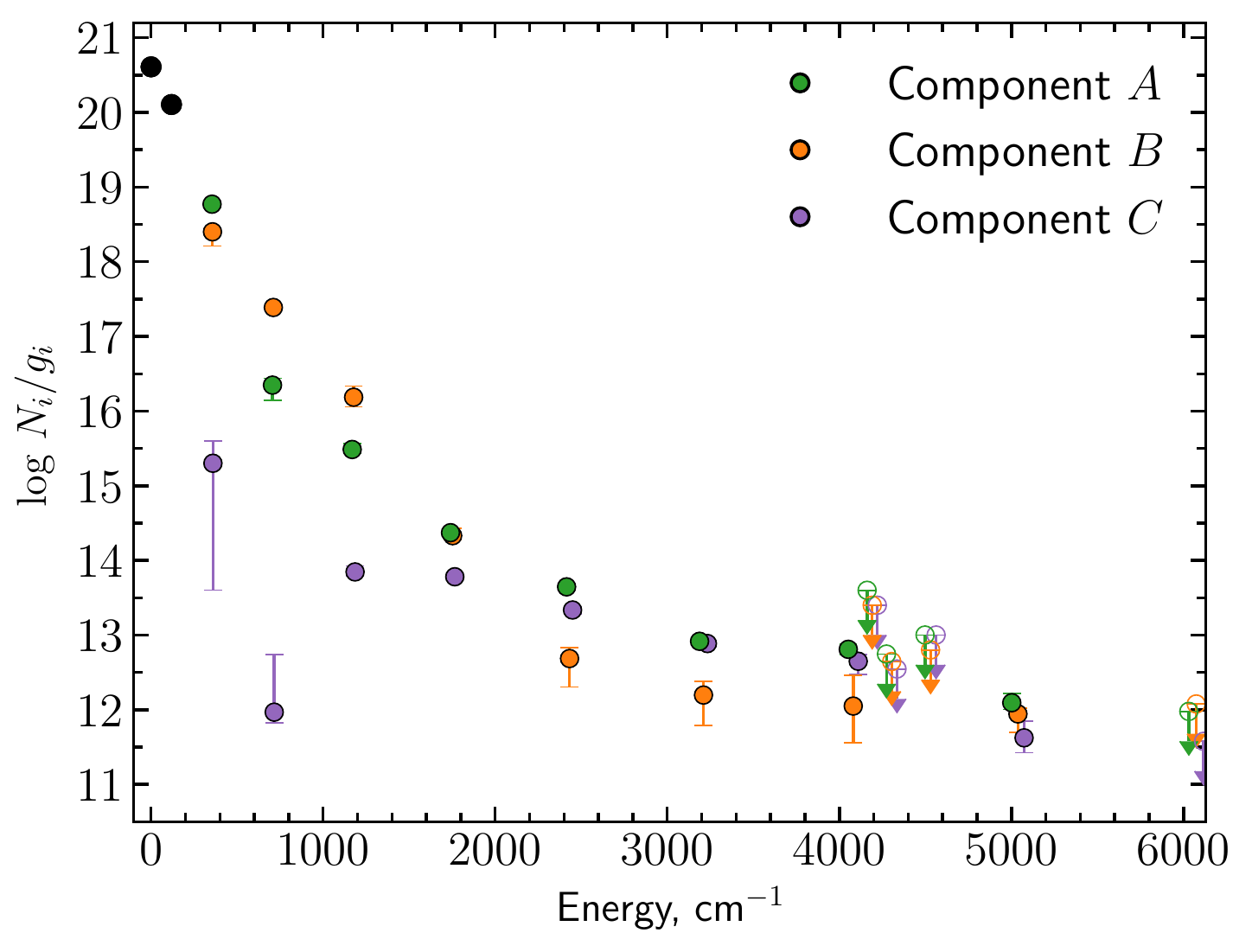}
		\caption{ Excitation diagrams of H$_2$ rotational levels for $A$ (green circles), $B$ (orange circles) and $C$ (violet circles) components of H$_2$ absorption system at z=2.786. The black points correspond to the $J=0$ and 1 levels, where we are only able to measure total column densities.}
		\label{H2Exc}
	\end{center}
\end{figure}

\label{lastpage}
\end{document}